\documentclass{aa}
\usepackage[pdftex]{graphicx}
\usepackage{longtable}
\usepackage{supertabular}
\def\be{\begin{equation}}
\def\ee{\end{equation}}
\usepackage{natbib}
\usepackage{amssymb}
\begin{document}
\title{The Cool-Core Bias in X-ray Galaxy Cluster Samples I: Method And Application To HIFLUGCS}
\author{D. Eckert\inst{1}, S. Molendi\inst{1} \& S. Paltani\inst{2}}
\institute{$^1$INAF/IASF-Milano, Via E. Bassini 15, 20133 Milano, Italy\\
$^2$ISDC Data Centre for Astrophysics, University of Geneva, 16, ch. d'Ecogia, 1290 Versoix, Switzerland}
\abstract{}{When selecting flux-limited cluster samples, the detection efficiency of X-ray instruments is not the same for centrally-peaked and flat objects, which introduces a bias in flux-limited cluster samples. We quantify this effect in the case of a well-known cluster sample, HIFLUGCS.}
{We simulate a population of X-ray clusters with various surface-brightness profiles, and use the instrumental characteristics of the \textit{ROSAT} All-Sky Survey (RASS) to select flux-limited samples similar to the HIFLUGCS sample and predict the expected bias. For comparison, we also estimate observationally the bias in the HIFLUGCS sample using \textit{XMM-Newton} and \textit{ROSAT} data.}
{We find that the selection of X-ray cluster samples is significantly biased ($\sim29\%$) in favor of the peaked, Cool-Core (CC) objects, with respect to Non-Cool-Core (NCC) systems. Interestingly, we find that the bias affects the low-mass, nearby objects (groups, poor clusters) much more than the more luminous objects (i.e massive clusters). We also note a moderate increase of the bias for the more distant systems.}
{Observationally, we propose to select the objects according to their flux in a well-defined physical range excluding the cores, $0.2r_{500}-r_{500}$, to get rid of the bias. From the fluxes in this range, we reject 13 clusters out of the 64 in the HIFLUGCS sample, none of which appears to be NCC. As a result, we estimate that less than half (35-37\%) of the galaxy clusters in the local Universe are strong CC. In the paradigm where the CC objects trace relaxed clusters as opposed to unrelaxed, merging objects, this implies that to the present day the majority of the objects are not in a relaxed state. From this result, we estimate a rate of heating events of $\sim1/3$ Gyr$^{-1}$ per dark-matter halo.}
\authorrunning{Eckert, D. et al.}
\titlerunning{The Cool-Core Bias in X-ray Galaxy Cluster Samples I}

\maketitle
\section{Introduction}

Galaxy clusters are the biggest gravitationally-bound structures in the Universe. They are filled with a hot ($kT\sim 1-10$ keV) plasma, the Intra-Cluster Medium (ICM), which has been heated to X-ray emitting temperatures by gravitational collapse. In the original ``cooling-flow" scenario \citep{fabian}, all clusters after relaxation from a major merging event should evolve into the cooling-flow state, where the gas condensates in the central regions and then cools through radiative processes until it eventually forms stars. This paradigm was supported by observations of prominent surface-brightness peaks and temperature drops in the cores of clusters, which were supposed to be associated with the central cooling flow. However, this picture was not confirmed by the latest generation of X-ray telescopes (\textit{XMM-Newton}, \textit{Chandra}), which found no spectroscopic evidence for the presence of the cooling gas in the central regions of clusters predicted by the cooling-flow model \citep{peterson,kaastra01}. These results lead to the revision of the classification of galaxy clusters, which were thereafter categorized into ``cool-core" (CC) and ``non-cool-core" (NCC) objects \citep{molendi01}. Since in the center of clusters the cooling time can be much shorter than the Hubble time, these results imply the existence of a heating mechanism which is responsible for quenching the cooling flow. Feedback from Active Galactic Nuclei (AGN) in the central galaxies is the most probable source of heating in the ICM \citep[e.g.,][]{mcnamara}.

Apart from the nature of the heating source, the failure of the cooling-flow model also has repercussions on the formation scenario of galaxy clusters. Indeed, some numerical studies of cluster mergers have found that cool cores are hardly disrupted by merging events \citep{poole}, and that the state of a cluster (CC or NCC) is determined once for all during the formation process \citep{mccarthy}. Recent observational works have shown that the population of clusters is bimodal \citep{cavagnolo}, which could support this scenario because of the lack of intermediate objects. However, other studies found no evidence for bimodality in the distribution of clusters \citep[e.g.,][]{pratt}, and the recent identification of regions reminiscent of cool cores in merging clusters (``cool-core remnants", \citet{rm09}), probably associated with ancient cool cores disrupted by merging events, argues against this idea. Thus, the question of the formation process of cool cores is still an open one.

An important clue to the understanding of the formation process of cool cores is the observed ratio between CC and NCC clusters in the local Universe, which depends strongly on the different formation scenarios. Indeed, in the old cooling-flow scenario it was believed that the majority of clusters (70-90\%) in the local Universe had a cooling flow \citep{peres}, which is expected if all clusters naturally evolve into the cooling-flow state. In this case, one might also expect a strong dependence of the CC fraction on the redshift. Conversely, if the state of a cluster is determined only by an initial entropy injection event as suggested by some simulations \citep{mccarthy04}, a very different behavior can be expected.

Observationally, to measure the fraction of CC objects it is crucial to use a sample which is as complete and as free of selection biases as possible. In this respect, the very different surface-brightness profiles observed in CC and NCC objects might introduce a bias in the selection of X-ray flux-limited samples. Indeed, CC systems exhibit a prominent surface-brightness peak, as opposed to NCC clusters which show flat emission profiles.  The diversity of surface-brightness profiles might introduce a bias in favor of CCs in X-ray flux-limited samples. In this paper, we investigate the impact of such a flux selection on the measurement of the CC over NCC ratio when computed using a well-studied, nearby cluster sample, HIFLUGCS, which was extracted from the \textit{ROSAT} All-Sky Survey (RASS). 

HIFLUGCS \citep{reip} is a complete flux-limited sample extracted from RASS data. It comprises 64 clusters with a flux higher than $2.0\times10^{-11}$ ergs cm$^{-2}$ $\mbox{s}^{-1}$ (0.1-2.4 keV band). According to the authors, it is to the present day the largest complete X-ray selected sample of galaxy clusters. Additionally, an extended sample of 106 objects also exists, which comprises objects which were originally excluded for completeness reasons (in particular because of high absorption). The objects of the extended sample are nearby ($z<0.2$) and span a luminosity range between $\sim10^{43}$ and $\sim10^{45}$ $h_{72}^{-2}$ ergs $\mbox{s}^{-1}$.

From \textit{ROSAT} and \textit{ASCA} data, \citet{chen} estimated the fraction of CC to be 49\% in the HIFLUGCS sample. More recently, studies have been carried out using \textit{Chandra} data \citep{mittal,hudson}, which clearly demonstrate the existence of a class of intermediate objects. In particular, \citet{hudson} (hereafter, H10) claim a fraction of 44\% ``strong CC" objects and 28\%  ``weak CC" clusters, where the classification of the objects was performed using the central cooling time (CCT), which was found to be the best indicator of the state of a cluster. Strong CC objects were defined as objects which exhibit a central cooling time $<1$ Gyr, while weak CC objects have 1 Gyr $<$ CCT $<$ 7.7 Gyr, and clusters with a CCT $>$ 7.7 Gyr are classified as NCC. In addition, H10 also analyzed the effect of the enhanced luminosity of cool cores on the number of CC objects using numerical simulations. The authors estimate that a true fraction of 31\% strong CC objects is sufficient to explain the value of 44\% measured in the HIFLUGCS sample. Comparing the HIFLUGCS clusters with a sample of high-redshift objects ($z>0.5$), \citet{vikhlinin} noted a decrease of the CC fraction with redshift, which they interpret as evidence for a higher rate of merging events. However, this result might have been affected by a bias against CCs \citep{santos}.

In the following, we present our analysis of the bias introduced in X-ray cluster samples by the sharp surface-brightness profiles of CC clusters (hereafter, CC bias), expanding on the work of H10, with the aim of providing a reliable measurement of the fraction of CC vs NCC clusters. The reader should note that the effect investigated here and the one analyzed by H10 are slightly different. While H10 analyzed the intrinsic effect induced by the increased luminosity of CC clusters, in this paper we investigate the bias introduced by the different detection efficiency of clusters for different surface-brightness profiles. For this work, we use both numerical simulations (Sect. \ref{sim}-\ref{sres}) and observations (Sect. \ref{data}-\ref{subsamp}). As a result, we compute the fraction of CC vs NCC objects in the local Universe taking into account the selection bias. In a companion paper, we use our simulation tool to compute the bias expected in several other well-known samples and make predictions for the future X-ray survey missions, eROSITA \citep{erosita} and WFXT \citep{wfxt}.

Throughout the paper, we assume a $\Lambda$CDM cosmology with $\Omega_m=0.27$, $\Omega_\Lambda=0.73$ and $H_0=72$ km $\mbox{s}^{-1}$ Mpc$^{-1}$.

\section{The simulation}
\label{sim}

To estimate the CC bias introduced into a sample, we use a Monte Carlo approach where clusters are distributed following the observed luminosity function, with a known input fraction of CC vs NCC objects, and then selected using the criteria of the corresponding sample. A realistic description of the survey properties (RASS) and of the instrumental characteristics are implemented to perform the selection. The fraction of CC vs NCC clusters is then computed from the selected sample and compared with the input ratio. The dependence of the bias on several input parameters (luminosity, redshift, absorption) is also analyzed.

\subsection{Input distributions of luminosity, redshift and column density}

The first step for the simulation is to select randomly a luminosity in the 0.1-2.4 keV band ($L_X$) and a redshift ($z$), following the observed distribution of clusters, i.e.
\be N_{cl}(L_X,z)=F(L_X,z)\left(\frac{dV}{dz}\right)dL_X\,dz,\label{ncl}\ee
where the X-ray Luminosity Function (XLF) $F(L_X,z)$ has the form
\be F(L_X,z)=C(z)L_X^{-\alpha}\exp(-L_X/L_\star(z)).\label{lfunc}\ee
The dependence of the luminosity function on the redshift is given by $L_\star(z)=L_\star(0)(1+z)^A$ and $C(z)=C(0)(1+z)^B$, where $C(0)$ is the normalization of the luminosity function at $z=0$ \citep{mullis}. For the parameters of the luminosity function we use the values extracted from the BCS survey: $\alpha=1.85$, $L_\star(0)=2.8\times 10^{44}$ $h_{72}^{-2}$ ergs $\mbox{s}^{-1}$ \citep{ebeling}. The cosmological evolution of the XLF follows the work of \citet{mullis}, $A=-2.3$ and $B=1.3$.

In the case of HIFLUGCS, we restrict ourselves to $L_X>2\times10^{42}$ ergs $\mbox{s}^{-1}$ and $z<0.25$ to cover all the parameter space of the sample. At low redshift, the cosmological dependence of the XLF is negligible, and therefore the redshift distribution goes like $z^2$, given that the number of objects is proportional to $\frac{dV}{dz}\sim z^2$. The luminosity distribution follows approximately a cut-off power law as expected from Eq. \ref{lfunc}. From the luminosity, we use the $L_X-T$ relation, 

\be L_X=A \left(\frac{kT_{vir}}{\mbox{6 keV}}\right)^\alpha, \ee

\noindent to get the corresponding temperature. For the parameters of the relation, we use the values derived by \citet{andersson}, $A=1.88\times10^{45}$ ergs $\mbox{s}^{-1}$ (2-10 keV band) and $\alpha=2.79$. We neglect the scatter of the relation. For galaxy groups, the relation is steeper \citep{helsdon}, so at low temperatures ($kT<1$ keV) we use the corresponding determination of the $L_X-T$ relation ($A=9.55\times10^{42}$ ergs s$^{-1}$, $\alpha=4.9$). For each value of $L_X$ and $z$, we derive the 0.1-2.4 keV flux of the corresponding object in the observer frame.

In addition to the luminosity and redshift, we also simulated a distribution of galactic hydrogen column density ($N_H$). For this purpose, we used the 21 cm maps from \citet{kalberla} and computed the distribution of $N_H$, restricting to galactic latitudes $|l|>20^\circ$ to match the considered samples. Values of $N_H$ were then simulated following the observed distribution.

\subsection{Input surface brightness profiles}
\label{params}

For the surface-brightness profiles of the simulated clusters, we use two different sets of parameters describing different physical cases. For simplicity, we divide the population of clusters strictly into CC and NCC clusters (i.e. we neglect intermediate objects). Namely, for NCC clusters we assume a surface-brightness profile described by a beta model \citep{cavaliere}, 
\be S(r)=S_0\left[1+(r/r_{c})^2\right]^{-3\beta+1/2},\label{beta}\ee
while for CC clusters we assume a double-beta model,
\begin{eqnarray} 
\nonumber S(r) & = & S_1\left(\left[1+(r/r_{c1})^2\right]^{-3\beta+1/2}\right. \\
 & & \left. +R\left[1+(r/r_{c2})^2\right]^{-3\beta+1/2}\right),\label{dbeta}\end{eqnarray}
where we use the same value of $\beta$ for the 2 beta components, and $R$ is the ratio between the two components at $r=0$. 

In the first case, we choose fixed values for the relevant parameters, which were estimated by averaging the observed quantities from the sample of \citet{mohr}. In both cases, we use a common $\beta$ value of 0.64. For NCC clusters, a core radius $r_c=230$ kpc was chosen. For CC clusters, we use a core radius for the smaller beta component, $r_{c2}$, of $40$ kpc, while the core radius of the broader beta component, $r_{c1}$, is set to 170 kpc, a smaller value than for NCC clusters. All these parameters show relatively small scatter in the \citet{mohr} sample. The most uncertain parameter of the simulation is the ratio between the 2 beta components at $r=0$ for CC clusters, namely $R$. From the \citet{mohr} sample, this quantity has a mean value of $\sim15$ with large scatter, the values ranging from 1.5 up to $\sim$100 from an object to another. In the following, we use a constant value $R=15$ for CC clusters. Indeed, we found that the results of the simulation do not differ significantly when a single value or a distribution of values is used for the relevant parameters (see Sect. \ref{sres}). For the remainder of the paper, we refer to this set of parameters as case I.

As an alternative approach, we also performed simulations in which the various core radii are defined as a fixed fraction of $r_{500}$\footnote{For a given over-density factor $\Delta$, $r_\Delta$ is defined as the radius within which the total mean density is a factor $\Delta$ above the critical density.}, hereafter case II. More specifically, for each simulated cluster we use the scaling relation of \citet{arnaud} to compute the expected value of $r_{500}$, and choose for the various core radii a fixed fraction of $r_{500}$. In terms of $r_{500}$, for NCC systems we use $r_c=0.19r_{500}$, while for CC objects we choose $r_{c1}=0.13r_{500}$ and $r_{c2}=0.03r_{500}$.

It is unclear which of the two parameter sets should provide a better description of the data. Indeed, the first approach assumes that in cluster cores feedback effects are dominant, thus creating universal core properties which do not depend on cluster mass. Conversely, the second approach neglects the feedback effects and assumes that the core properties are only determined by gravitational processes, in which case they should follow a self-similar relation. Since it is known that both effects are important, our two choices of scale radii should provide limiting cases.

\subsection{ROSAT specifics and source detection}

The \textit{ROSAT} all-sky survey (RASS) covered the whole sky with a typical exposure time of a few hundred seconds \citep{cruddace}. The instrument featured an effective area $A\sim400$ cm$^2$ @ 1 keV and a spatial resolution of $\sim$ 20 arcsec FWHM averaged over the FOV. The background count rate, which we estimated by computing the mean number of counts in RASS images, is low, $\sim10^{-3}$ counts $\mbox{s}^{-1}$ arcmin$^{-2}$ (0.1-2.4 keV). To convert fluxes to RASS count rates, we used the XSPEC package \citep{xspec} v12.6.0 and the \textit{ROSAT}/PSPC response. We assumed that the spectrum of each source is described by an absorbed MEKAL model \citep{kaastra} with varying temperature and hydrogen column density. The conversion factor from flux to count rate was then computed for a range of values of $kT$ and $N_H$. For any initial value of temperature and $N_H$, we estimate the conversion factor by interpolating the nearest points.

In the case of extended sources, the choice of the integration radius from the surface-brightness peak is crucial. Beyond a given radius, the surface brightness of the source becomes smaller than that of the background, and therefore the \emph{observed} flux of a cluster in a real observation is always smaller than the total flux of the cluster. For this reason, the fluxes computed as described above cannot reproduce accurately the RASS.

To overcome this difficulty, for each cluster with a given total flux $F$ we define a radius $r_{max}$ such that
\be F_{obs}=\int_0^{r_{max}} S(r)\,2\pi r\, dr.\label{fobs}\ee
Obviously, the surface-brightness profile $S(r)$ is normalized in such a way that
\be \int_0^\infty S(r)\,2\pi r\, dr = F.\ee
For a cluster to match the selection criterion of HIFLUGCS we impose that $F_{obs}$, not $F$, be above the flux limit of the sample (i.e. $2.0\times10^{-11}$ ergs cm$^{-2}$ $\mbox{s}^{-1}$).

\begin{figure}
\resizebox{\hsize}{!}{\includegraphics{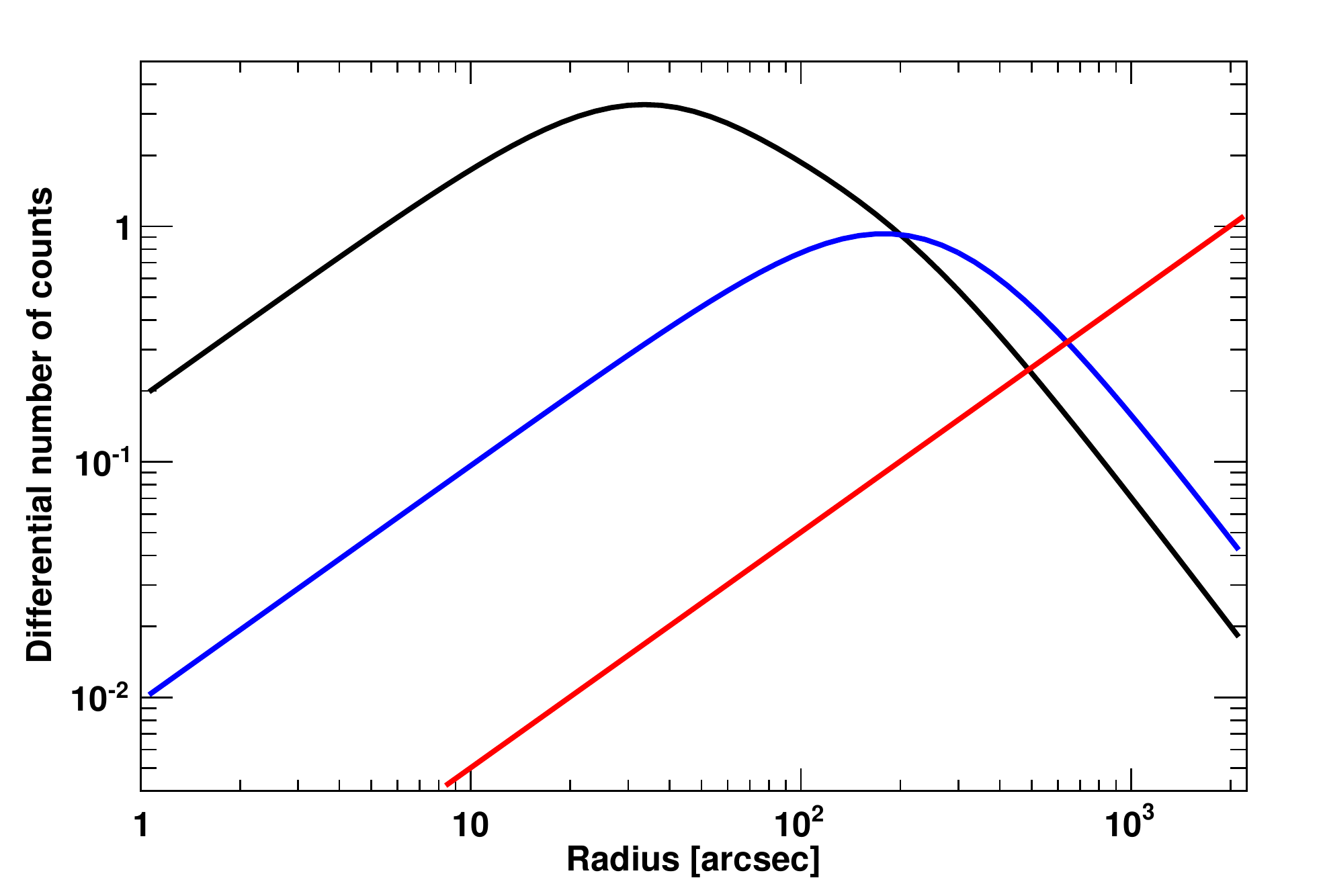}}
\caption{Differential number-of-counts profiles (in units of counts per arcmin) for a CC (black) and a NCC cluster (blue) with the same redshift ($z=0.05$) and luminosity ($L_X=10^{44}$ ergs $\mbox{s}^{-1}$), compared to the \textit{ROSAT}/PSPC background profile (red).}
\label{functions}
\end{figure}

Clearly, a realistic definition of $r_{max}$ is important to reproduce accurately the observed samples. Given that the original fluxes were computed inside the maximum radius up to which the source is detected \citep{reip}, we chose to use for $r_{max}$ the radius for which the surface-brightness of the source is equal to that of the background. This should be regarded as a conservative choice, since in most cases the source contribution can be measured up to a somewhat larger radius. In Fig. \ref{functions} we show the differential number-of-counts profiles, i.e. the surface brightness times $2\pi r$, for the \textit{ROSAT}/PSPC background and two clusters, a CC and a NCC, both at $z=0.05$ and with a luminosity of $10^{44}$ ergs s$^{-1}$ (total flux $F\sim2\times10^{-11}$ ergs cm$^{-2}$ $\mbox{s}^{-1}$). The value of $r_{max}$ in the two cases is given by the intersection between the black/blue curve and the red curve. In the specific case shown here, $r_{max}=490^{\prime\prime}$ for the CC case and $r_{max}=640^{\prime\prime}$ for the NCC case. In terms of flux, this results in a $\sim25\%$ higher observed flux for the CC cluster compared to the NCC cluster, for the same total flux.

\subsection{Construction of a simulated sample}

The critical parameter for the simulation, $c$, is the fraction of clusters hosting a cool core. Since we expect that the \emph{observed} value of $c$ in the samples is biased, we wish to compute the \emph{input} value of $c$ for which the fraction of CC clusters in a simulated sample reproduces the \emph{observed} value in a given observed sample. Therefore, for each simulated cluster we randomly choose whether it is CC or NCC, imposing a fraction $c_{inp}$ of CC clusters. We then compute the corresponding integration radius $r_{max}$ as explained above and calculate the \emph{observed} flux using Eq. \ref{fobs}. For a cluster to match the selection criterion of a sample, we then impose the corresponding flux limit. In addition, we also check the detection level of each source. More specifically, we impose that the signal-to-noise of a source, defined as 

\be S/N=\frac{N_{source}}{\sqrt{N_{source}+N_{bkg}}},\ee

\noindent where $N_{source}$ and $N_{bkg}$ are the total number of counts of the source and the background integrated up to $r_{max}$, be above 5.0 to be selected in the sample. In the case of HIFLUGCS, all the objects are detected with $S/N>30$, so this condition is unnecessary, but in general this is not true.

Once the simulated sample is selected, we infer the fraction of CC clusters among the selected objects ($c_{obs}$), and compare this number with the input fraction $c_{inp}$.

\section{Simulation results}
\label{sres}

We started by simulating a population according to case I (see Sect. \ref{params}). In a simulation of $10^6$ objects, we found that $\sim900$ match the selection criteria, which allows us to compute the observed fraction $c_{obs}$ with good accuracy. We then performed simulations for a range of values of $c$ and found that the observed value of 49\% \citep{chen} corresponds to an input ratio $c_{inp}=0.38\pm0.02$, i.e. to a significant bias of $\sim$29\%. In case II (see Sect. \ref{params}), we find a similar result ($c_{inp}=0.37\pm0.02$). We also investigated how the bias changes when we use distributions of the relevant parameters rather than fixed values. The results are shown in Table \ref{tabdist}, using the same input value $c_{inp}=0.38$. Introducing distributions of parameters has little influence on the output results, so for the remaining of the paper we use fixed values for the surface-brightness parameters.

\begin{table}
\caption{\label{tabdist}Results of different simulations of $10^6$ objects with an input CC fraction $c_{inp}=0.38$. Left: Surface-brightness parameters used for the simulation (see Sect. \ref{params}). Right: Observed fraction of CC clusters in the simulated samples.}
\begin{center}
\begin{tabular}{cc}
\hline
\hline
Simulation & $c_{obs}$\\
\hline
All fixed & $0.49\pm0.02$\\
Distribution of $\beta$ & $0.47\pm0.02$\\
Distribution of $r_c, r_{c1} $ and $r_{c2}$ & $0.50\pm0.02$\\
Distribution of $R$ & $0.51\pm0.02$\\
\end{tabular}
\end{center}
\end{table}

\begin{figure}
\resizebox{\hsize}{!}{\hbox{\includegraphics[height=3cm]{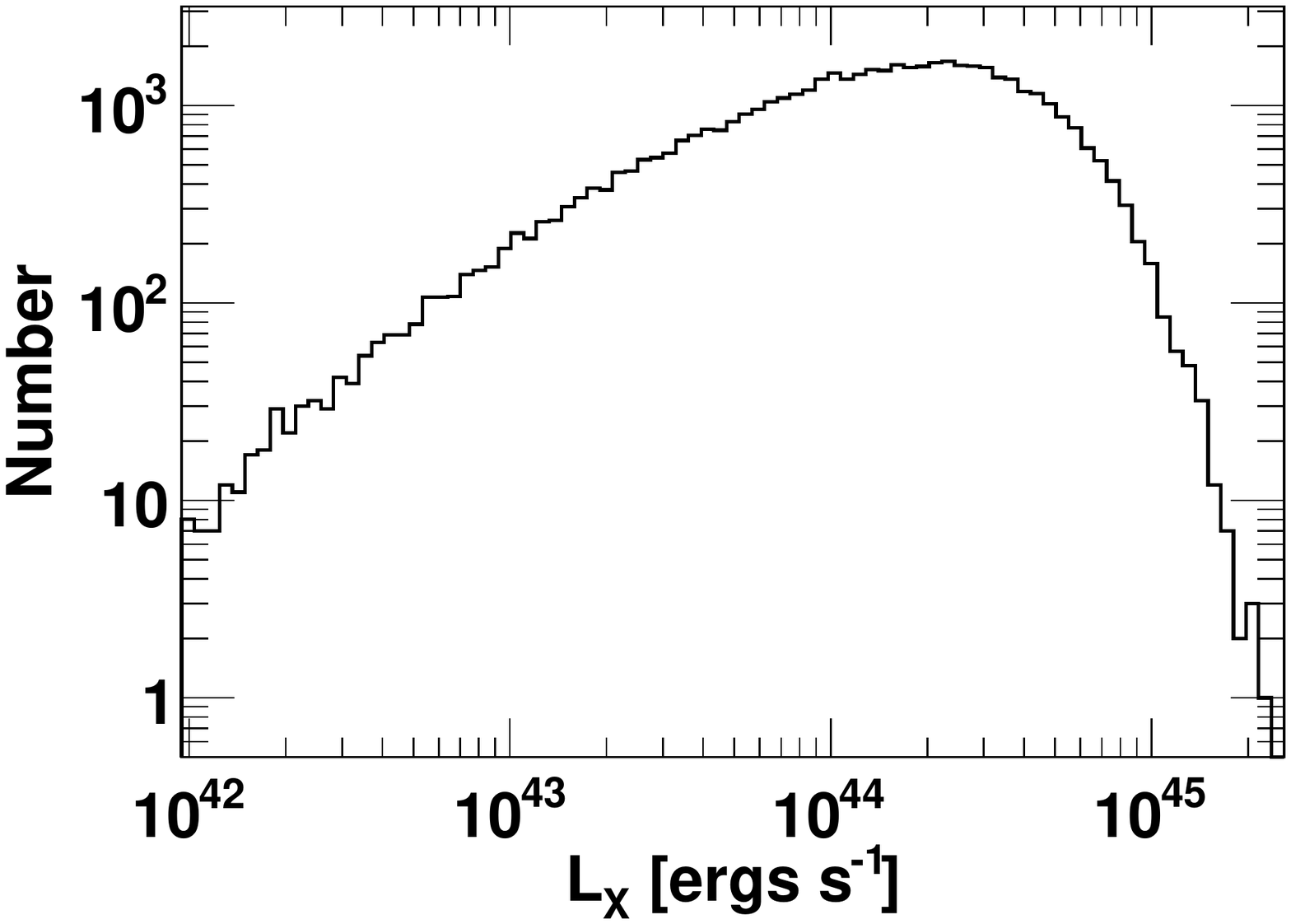}\includegraphics[height=3.05cm]{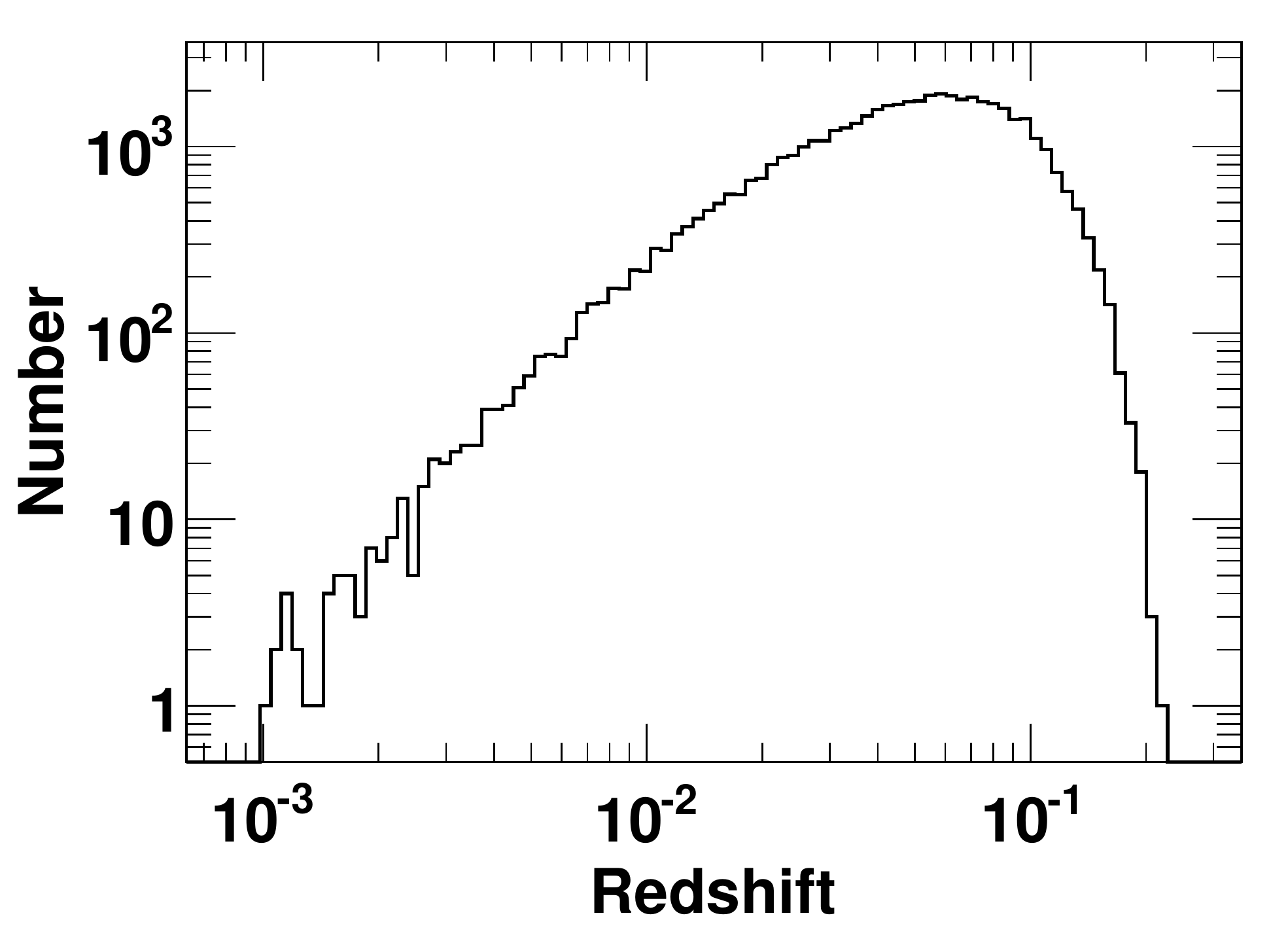}}}
\caption{Distribution of luminosities (left) and redshifts (right) of the objects in the simulated flux-limited sample.}
\label{selected}
\end{figure}

In addition to the total bias, it is also interesting to investigate the dependence of $c_{obs}$ on other quantities, in particular on luminosity, redshift and hydrogen column density. A simulation with a very large number of objects is required to study the dependence of $c_{obs}$ on these parameters. Fixing the value of $c_{inp}$ to 0.38, we then performed large simulations of $10^8$ objects for the two parameter sets (case I and case II), where as expected $\sim90,000$ match the selection criterion. Figure \ref{selected} shows the luminosity and redshift distributions of the selected objects. The distributions peak at a luminosity $\sim3\times10^{44}$ ergs $\mbox{s}^{-1}$ and a redshift $\sim0.05$, which matches well the observed quantities of the HIFLUGCS sample \citep{reip}. Moreover, as expected, the log$N$-log$S$ of the sample is well-represented by a power law with an index $-3/2$, in agreement with the properties of the actual sample.

\begin{figure*}
\resizebox{\hsize}{!}{\hbox{\includegraphics{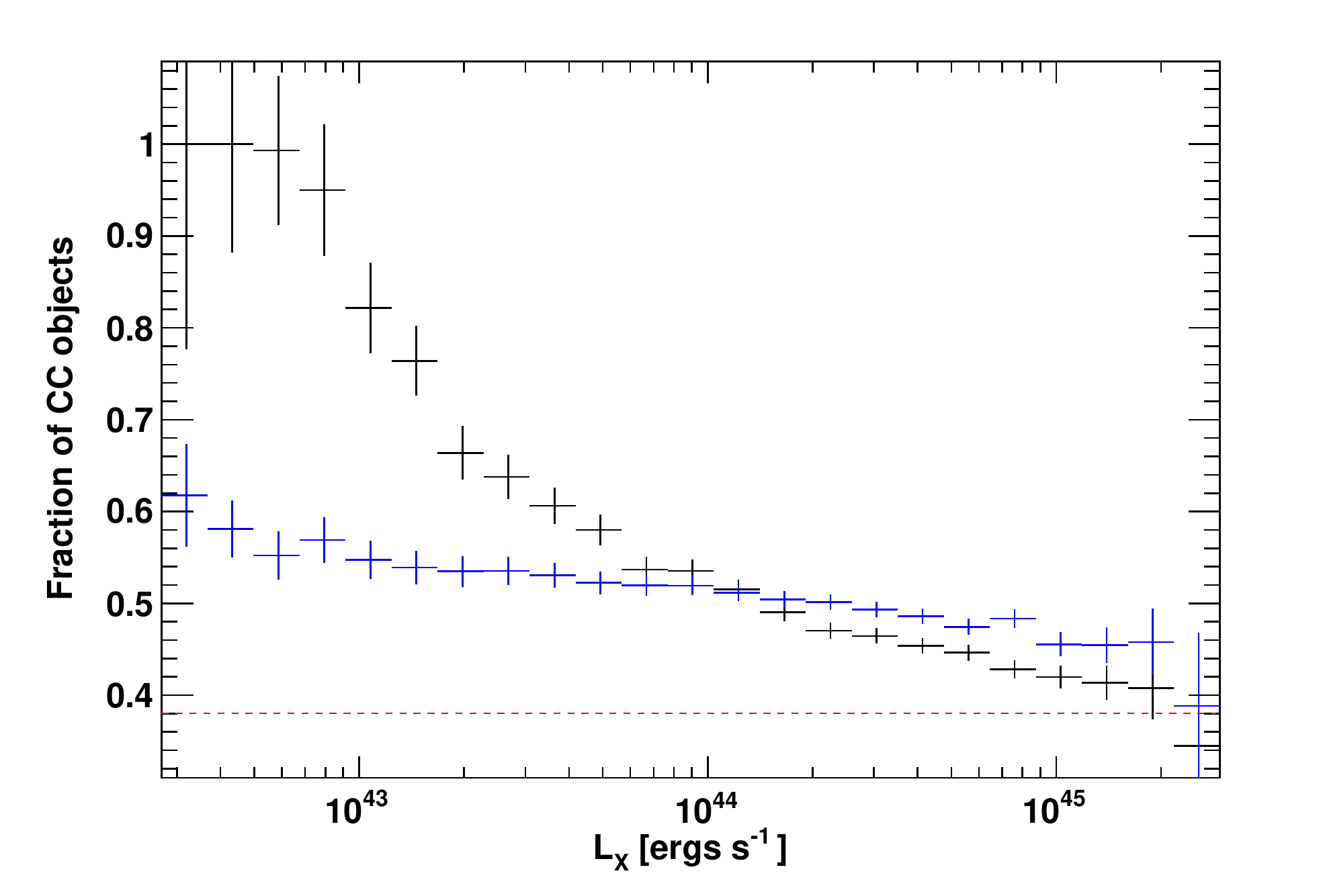}\includegraphics{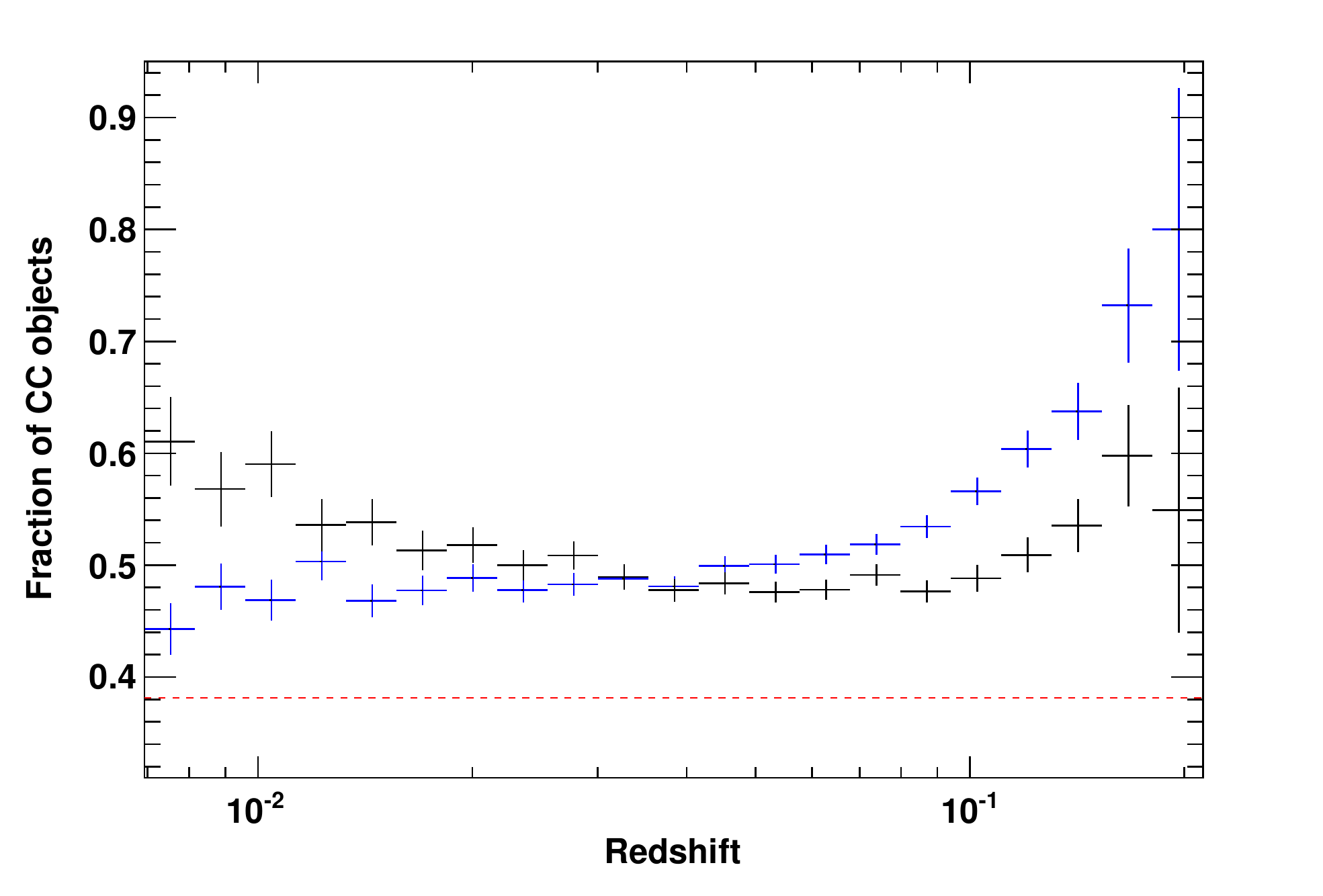}}}
\caption{Dependence of the observed fraction of CC objects, $c_{obs}$, on the 0.1-2.4 keV luminosity (left) and the redshift (right). In both cases, the black or blue data show the simulation results for case I or case II, respectively (see text). The dashed red lines show the input fraction of CC objects, $c_{inp}$, which was fixed to 0.38 for the simulations.}
\label{ccfrac}
\end{figure*}

We studied the dependence of $c_{obs}$ on redshift, luminosity and absorption. The resulting plots are shown in Fig. \ref{ccfrac} (black: case I; blue: case II). Since they represent extreme cases where the core radii are determined only by gravitation (blue) and feedback (black) effects, the two curves should give lower and upper bounds to the actual values. In case I, we observe a strong dependence of $c_{obs}$ on the luminosity. In particular, the low-luminosity objects ($L_X<10^{43}$ ergs $\mbox{s}^{-1}$) are very strongly biased towards CC clusters. Since the number of objects in this luminosity range is relatively small (see the left panel of Fig. \ref{selected}), this effect does not have a big influence on the mean value of $c_{obs}$. This effect is related to the background level. Indeed, these objects are both low-luminosity and nearby, so their surface-brightness peak is small. In the case of some NCC clusters, it can even be below the surface brightness of the background, in which case $r_{max}=0$ and the source is not detected at all. Conversely, because of their peaked profiles CC objects with the same total luminosity are easily detected. Therefore, CC clusters completely dominate the detected objects at luminosities below $10^{43}$ ergs s$^{-1}$. In case II, this effect, albeit present, is much less severe, because low-luminosity (i.e. low-mass) systems have smaller core radii. As a result, they are more concentrated, and the effect becomes less important. It has been claimed \citep[e.g.,][]{chen} that low-mass systems (groups, poor clusters) are predominantly CC. If the CC bias is not taken into account, we see from our simulation that one would immediately reach such a conclusion, even if the true fraction of CC objects would not depend at all on luminosity.

From Fig. \ref{ccfrac} we also observe a dependence of $c_{obs}$ on the redshift of the object. In particular, we note a significant increase in $c_{obs}$ at the highest redshifts. This is a consequence of the cut-off in the luminosity function. At rather high redshifts, only the most luminous objects are selected, and above the cut-off luminosity their number decreases quickly with increasing luminosity. Since, for a given limiting flux, the associated luminosity is smaller for CC than for NCC systems, the relative number of CC objects will be significantly larger. We note that unlike the low-luminosity effect, which is due to a combination of the intrinsic properties of the different population and of the background level, this effect is caused by the flux limit. With a lower flux limit, this effect would not disappear, but would be shifted to higher redshift, where the effect of the cut-off in the luminosity function will start to appear. Unlike the luminosity dependence, in case II we find a stronger evolution of $c_{obs}$. In this case, since we are only selecting high-mass systems the core radii are proportionally larger, which increases the bias.

As expected, the selected sample shows very little dependence on $N_H$. Indeed, the measured value of $c_{obs}$ is constant for varying $N_H$, and the total fraction of CC is very similar to the one obtained when the effect of absorption is completely neglected.

\begin{figure*}
\resizebox{\hsize}{!}{\hbox{\includegraphics{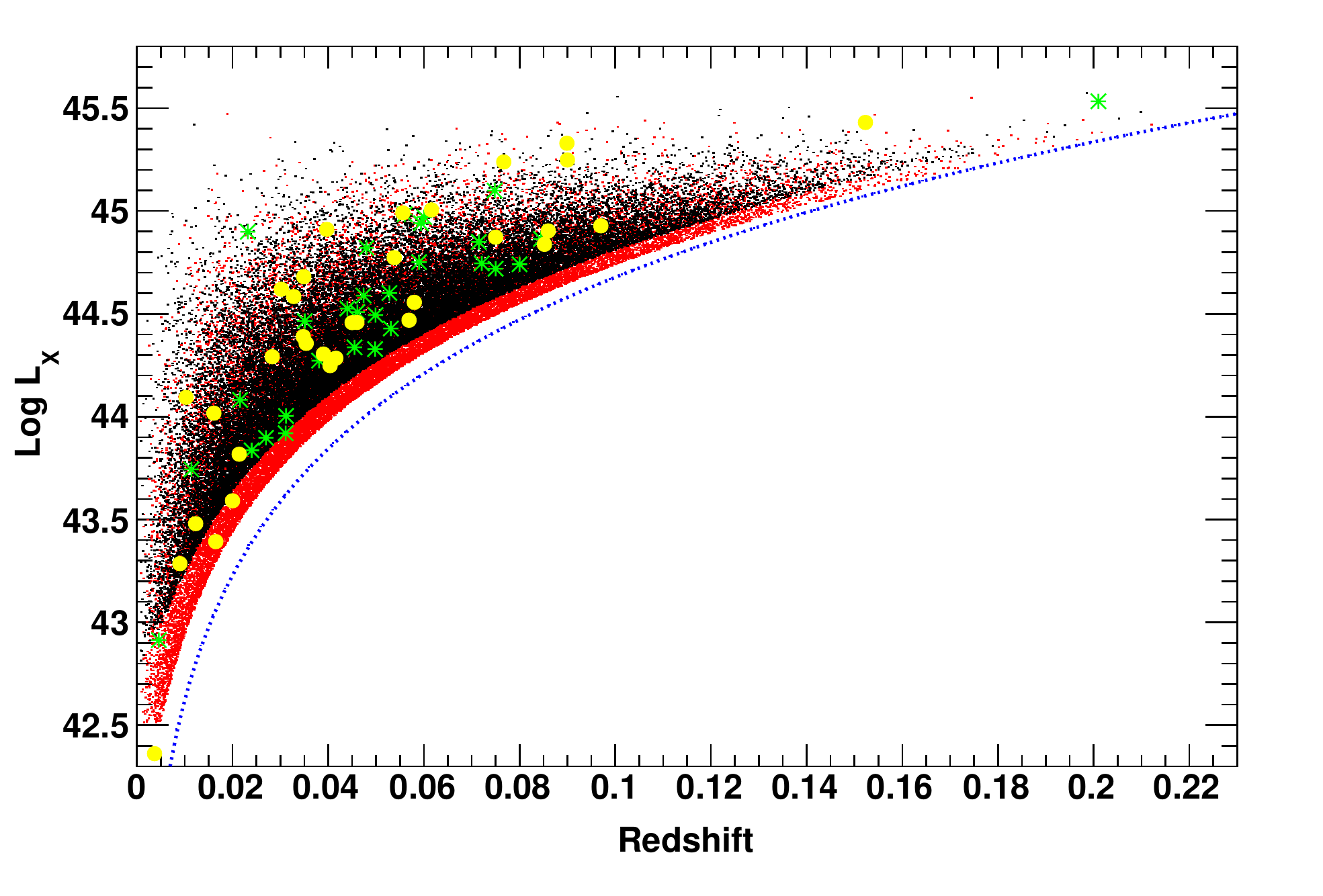}}}
\caption{Comparison between simulated objects and observational data in the $\log L_X$ vs $z$ plane. The black and red dots represent the simulated NCC and CC clusters, respectively. The green symbols (NCC) and yellow circles (CC) show the data from the extended HIFLUGCS sample \citep{chen}. The dashed blue line shows the flux limit for point sources.}
\label{lxz}
\end{figure*}

To visualize the simulated objects and compare them with the actual data, we plotted all the detected clusters in the $L_X-z$ plane, and added to the plot the data of the HIFLUGCS sample from \citet{chen}. In this case, we classified the clusters as CC/NCC on the basis of their classical mass deposition rate, CC clusters being the ones showing a non-zero mass deposition rate. The resulting plot for case I can be seen in Fig. \ref{lxz}. Black (NCC) and red (CC) dots each represent a single simulated cluster, while the green symbols (NCC) and yellow circles (CC) show the actual data for the objects of the extended HIFLUGCS sample.

Interestingly, we see in Fig. \ref{lxz} that at all redshifts the simulated CC clusters populate a broader range of luminosities than NCC clusters. In other terms, the total flux limit of the sample is different for CC and NCC clusters. At low luminosities ($\log L_X<43.5$) CC objects dominate the observed and simulated data, which again indicates an observational bias (see the left panel of Fig. \ref{ccfrac}). All but one NCC clusters lie in the area populated by the black dots, while the remaining object is found just below the limit. This indicates that our simulation reproduces well the properties of the sample. A similar conclusion is reached for case II.

\section{XMM-Newton and ROSAT analysis of surface-brightness profiles from the HIFLUGCS sample}
\label{data}

We analyzed available data for the HIFLUGCS sample, with the aim of providing an alternative estimate of the CC bias and of extracting a subsample that is unaffected by the CC bias. We proceeded with the systematic analysis of surface-brightness profiles for all clusters in the HIFLUGCS sample. Since the sample is mostly composed of nearby objects, the Field Of View (FOV) of the selected instruments is a key feature to observe regions as large as possible. Indeed, we are interested in the extraction of total fluxes within a given physical radius, so a detailed characterization of the central regions is not necessary, but it is important to detect the emission to relatively large radii. As a result, we decided to use \textit{XMM-Newton}/EPIC and \textit{ROSAT}/PSPC pointed data to pursue our goals, since they are the X-ray instruments with the largest FOV. In most cases, when both \textit{XMM-Newton} and \textit{ROSAT} pointed data were available, we combined the surface-brightness profiles from the two instruments to take advantage of the better \textit{XMM-Newton} point spread function (6 arcsec FWHM) in the central regions and of the larger FOV (1 square degree) and lower background of \textit{ROSAT} in the external regions. For very nearby objects ($z<0.015$), the \textit{ROSAT} PSF is sufficient to resolve the cores, and a detection to large radii is needed to constrain the parameters, so we restricted our analysis to \textit{ROSAT}. Conversely, for the more distant objects ($z>0.08$) the apparent size of the objects is smaller, so a better PSF is necessary to resolve the core. In these cases, only \textit{XMM-Newton} was used.

To construct a subsample free of selection bias, we measure fluxes in an annulus around the core to exclude the core component. We then perform a new selection on the basis of these fluxes. Generally, the flux of a cluster is not a well-defined quantity, since it is computed as the total flux within the area where the source is detected. However, in this work we wish to extract fluxes within a well-defined physical region, which we choose to be $r_{500}$. Since in most cases we do not detect the sources at this radius, a best-fit surface-brightness model is used to extrapolate the fluxes to $r_{500}$. We use the scaling relations of \citet{arnaud} and cluster virial temperatures from H10 to estimate $r_{500}$ for each object. Overall, this corresponds to a size which ranges from $\sim$ 500 kpc for small, group-like objects up to $\sim$ 1,500 kpc for the most massive, hot clusters. The values (physical and apparent) of $r_{500}$, the redshift and virial temperatures of all HIFLUGCS objects are summarized in Table \ref{tabobs}, together with the log of the available data.

\subsection{Data analysis}

\subsubsection{XMM-Newton}

\textit{XMM-Newton}/EPIC data analysis was carried out in a systematic way using the standard XMMSAS software v9.0. For each EPIC detector, light curves in the soft (2-5 keV) and hard band (10-12 keV) were extracted, and events were filtered to exclude flares. To minimize the instrumental background, we extracted images from the cleaned event lists in a narrow band \citep[0.7-1.2 keV,][]{ettori10}, which excludes the prominent Al ($\sim$ 1.5 keV)  and Si ($\sim$ 1.8 keV) background lines \citep{lm08}. As a result, the total background is dominated by the sky background, which is convolved with the vignetting of the instrument in the same way as the source signal. For this work, since we are interested in the characterization of the total flux it is important to detect the sources to large radii. Given that the objects considered here are bright, combining the three EPIC detectors is not necessary. We chose to use MOS because of its better imaging capabilities, and for simplicity we restricted our analysis to MOS2. To extract surface-brightness profiles, we generated MOS2 images with 4 arcsec spatial binning in the 0.7-1.2 keV band, and the corresponding exposure maps using the SAS tool \textrm{eexpmap}.

\subsubsection{ROSAT}

When available, we restricted our analysis to \textit{ROSAT}/PSPC pointed observations. We used the filtered event files available in the HEASARC archive\footnote{http://heasarc.gsfc.nasa.gov/} and extracted images using the XSELECT tool available within the HEASOFT\footnote{http://heasarc.nasa.gov/lheasoft/} software package with 15 arcsec bins in the 0.4-2.0 keV bands. Exposure maps in the same energy band were generated using the PSPC detector maps and the instrument attitude files through the \textrm{pcexpmap} tool. 

\subsection{Construction of surface-brightness profiles and flux measurements}
\label{fluxes}

For every cluster and instrument, we used the output count images and corresponding exposure maps to extract surface-brightness profiles. After the detection of individual point sources using a local background method, we excised from the images the corresponding regions. We then extracted total count profiles,  centered on the image centroid, defined as
\be (\langle x\rangle,\langle y\rangle)=\left(\frac{\sum x\cdot I_{x,y}}{\sum I_{x,y}},\frac{\sum y\cdot I_{x,y}}{\sum I_{x,y}} \right),\ee
 
\noindent where $I_{x,y}$ is the count rate in the pixel with image coordinates $(x,y)$.  The minimum bin size of the profile was set to 7 arcsec (\textit{XMM-Newton}) and 20 arcsec (\textit{ROSAT}), and the bins were grouped to ensure a minimum of 50 counts per bin and permit the use of $\chi^2$ statistics. The count profiles were then divided by the corresponding exposure to correct for the instrument vignetting. 

As a result, we obtained exposure-corrected profiles, which we fitted by a source + background model. A double-beta profile was preferred to the standard beta profile when the improvement to the fit was found to be statistically significant by the F-test. The background-subtracted source profile $SB(r)$ was then inferred. To ensure that the source dominates over the background, this procedure was applied up to the radius $r_{max}$ where the model source intensity is at least 2 times that of the background. The best-fit surface-brightness model was then used to extrapolate the fluxes to $r_{500}$. We define the total fluxes within $r_{500}$ as:

\be F_{tot}=\sum_{r<r_{max}} SB(r)\,w(r) + \int_{r_{max}}^{r_{500}} S(r)\,2\pi r\, dr, \label{totflux}\ee

\noindent where $SB(r)$ represents the measured background-subtracted surface-brightness profile in the bin corresponding to the radius $r$, $w(r)$ is the width of the bin, and $S(r)$ is the best-fit surface-brightness model. In the radial range where we use the actual data, the error on the measured flux is simply calculated by propagating the Poisson statistics. In the outer regions where we use the model surface-brightness profile to estimate the flux up to $r_{500}$, we estimate the error through a Monte Carlo approach: randomly-picking the model parameters, we compute the corresponding flux for each parameter set, and extract the resulting flux distribution. Since the core contribution is expected to be negligible in this range, when a double-beta model is required we fix the parameters of the inner beta component. Moreover, the outer core radius $r_{c1}$ and the $\beta$ parameters are strongly degenerate, so we only let $\beta$ vary within its uncertainties. The error on the second term of Eq. \ref{totflux} is then given by the RMS of the resulting flux distribution. Overall, in the vast majority of cases the statistical uncertainties on the extracted fluxes are small ($F/\Delta F\sim 50$), however in some cases where the parameters are poorly constrained the relative uncertainty on the flux measurement from 0 to $r_{500}$ can be quite large. 

\begin{figure}
\resizebox{\hsize}{!}{\includegraphics{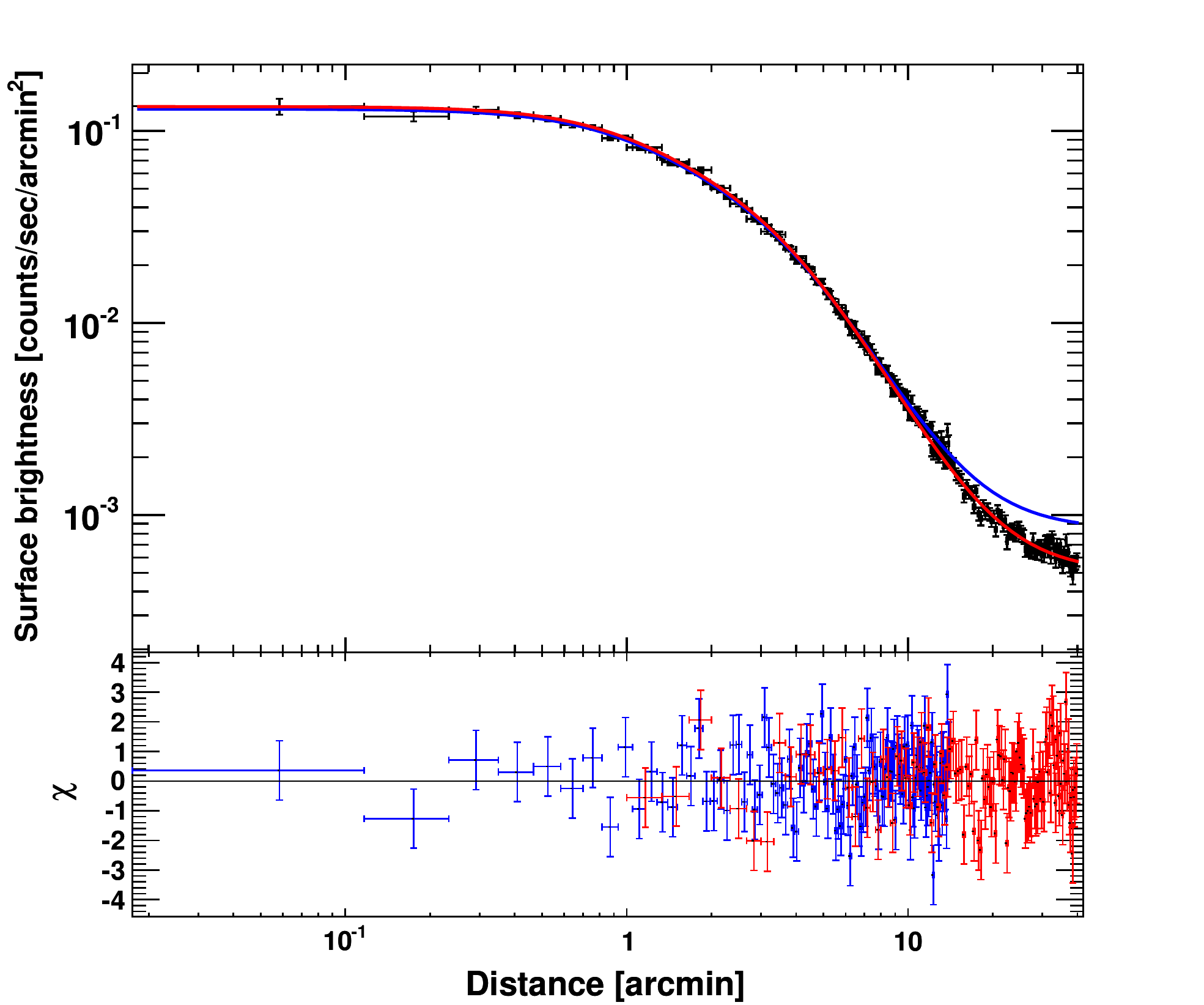}}
\caption{\textit{XMM-Newton}/MOS2 (blue) and \textit{ROSAT}/PSPC (red) surface-brightness profile of the cluster A3571, fitted jointly with a double-beta model. The individual normalizations and background levels were fitted independently. The higher \textit{XMM-Newton} data points and model at large radii are caused by a higher background. The bottom panel shows the deviations (in $\sigma$) from the surface-brightness model.}
\label{a3571}
\end{figure}

In a majority of cases (38 objects out of 64), we combined \textit{ROSAT} and \textit{XMM-Newton} surface-brightness profiles and fitted them jointly to get better constraints on the parameters. As an example, Fig. \ref{a3571} shows the \textit{XMM-Newton}/MOS2 (blue) and \textit{ROSAT}/PSPC (red) surface-brightness profiles of the cluster A3571 fitted by a double-beta profile plus a constant for the background. While the physical parameters of the model where fitted jointly, the individual normalizations and background levels where adjusted independently. In the innermost 1 arcmin of the profile, \textit{ROSAT} data were ignored to avoid problems of mixing caused by the broader PSF.

To convert the count rates into physical fluxes, we used the \textit{XMM-Newton}/MOS2 and \textit{ROSAT}/PSPC redistribution matrices and on-axis effective area, and used XSPEC to fold the spectrum of each source, modeled by an absorbed MEKAL model with the appropriate temperature and $N_H$, with the appropriate instrument response. The folded spectra were then used to convert the extracted count rates into unabsorbed physical fluxes in the 0.5-2.0 keV band. 

As a byproduct of this work, we were able to estimate the cross-calibration between the two instruments (see Appendix \ref{calib} for details). Indeed, since they are bright, persistent sources, clusters of galaxies are ideal sources to cross-calibrate between two independent missions. On average, we find that MOS2 gives $\sim$15\% higher fluxes compared to PSPC, and hence there is a clear cross-calibration issue which must be taken into account. Since the HIFLUGCS sample was selected using fluxes extracted from \textit{ROSAT} only, we decided to use the \textit{ROSAT} normalization to extract fluxes, and apply a 15\% cross-calibration factor to the \textit{XMM-Newton} fluxes. After applying such a renormalization, \textit{XMM-Newton} and \textit{ROSAT} fluxes agree with $\sim$5\% scatter.

\begin{figure}
\resizebox{\hsize}{!}{\includegraphics{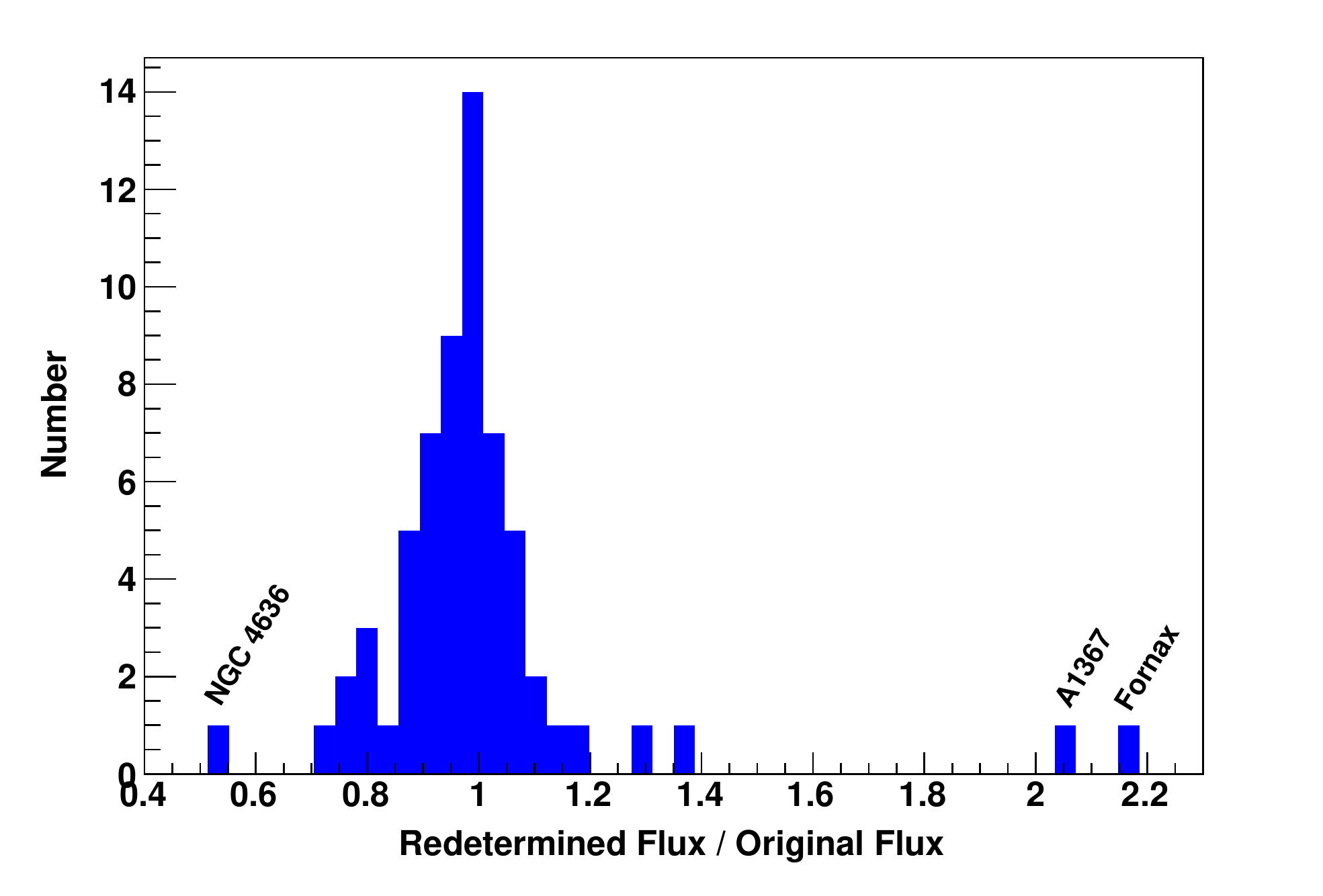}}
\caption{Distribution of the ratio between our redetermined fluxes and the original fluxes from \citet{reip}. Since the original fluxes were computed in the 0.1-2.4 keV band, they were rescaled to match our choice of energy band (0.5-2.0 keV). The most prominent outliers are highlighted. }
\label{fnew}
\end{figure}

The best-fit parameters and 1-$\sigma$ errors, as well as the total fluxes from \textit{XMM-Newton} and \textit{ROSAT} are summarized in Table \ref{tabpar}. The flux limit of the HIFLUGCS sample ($2.0\times10^{-11}$ ergs cm$^{-2}$ $\mbox{s}^{-1}$  in the 0.1-2.4 keV band) translates into a total flux of $1.2\times10^{-11}$ ergs cm$^{-2}$ $\mbox{s}^{-1}$ in the 0.5-2.0 keV band for a typical temperature of 4 keV. All \textit{XMM-Newton} fluxes should be rescaled by 15\% when compared to \emph{ROSAT} to account for the systematic calibration difference. Figure $\ref{fnew}$ shows the distribution of ratio between the \textit{ROSAT} fluxes extracted following Eq. \ref{totflux} and the original fluxes quoted by \citet{reip}. In the vast majority of cases, we can see that the fluxes estimated using Eq. \ref{totflux} agree very well with the original fluxes. In three cases, our fluxes differ from the original fluxes by a factor $\sim2$, either higher (Fornax and A1367) or lower (NGC 4636). All of these are nearby, very extended objects. The difference can be explained by the low surface brightness of the outer regions, which prevents the detection of a large part of the flux, and by the instability of the extrapolation over a large radial range. Overall, this is however not a problem, given that our definition of cluster fluxes differs from the original method, where the fluxes were measured by integrating the flux within the area where the clusters are detected.

According to our analysis, one cluster (Zw III 54) does not have a sufficient total flux to match the selection criterion of the sample. This cluster is the weakest of the sample and was at the extreme limit of the original selection, and therefore, even though our redetermined flux does not differ significantly from the original one, we find its flux to be below the flux cut.

\section{The unbiased subsample}
\label{subsamp}

\subsection{Selection of the subsample}

To measure the CC fraction, we wish to extract a subsample of HIFLUGCS which should be free of the CC bias. The fraction of CC objects would then be computed from the subsample and not from HIFLUGCS. At variance with the cores, it has been found that in cluster outskirts the surface-brightness profiles are essentially self-similar \citep{neumann2,croston, lrm09}, so the selection based on the fluxes in the outer regions should be unbiased. We define a new sample by selecting those objects for which the flux in an annulus excluding the core, and not the total flux, is above a given flux limit. While the choice of the outer radius is straightforward (we select $r_{500}$), the same is not true for the inner radius, $r_{in}$. Here we require a value such that the core emission would be excluded as much as possible, but that would still allow a statistically accurate determination of the fluxes. To this end, we analyzed the distributions of the best-fit parameters from Table \ref{tabpar}. In the cases where a double-beta model was required by the fit, we searched for a typical radius at which the contribution of the core can be considered to be negligible. To do this, we looked for the radius above which the flux can be attributed mostly to the outer beta component. All in all, we find that $r_{in}=0.2r_{500}$ is a reasonable choice. With such a choice, in average 97\% of the flux is attributed to the main beta component, and the fluxes can still be computed with good accuracy. For comparison, this radius is larger than or comparable to the cooling radius $r_{cool}$ for which the cooling time exceeds 7.7 Gyr (H10). \citet{maughan} excised the region within $0.15 r_{500}$ to exclude the effects of the core. In this case, in average $\sim6\%$ of core flux is still included. For the remaining of the paper, we compute the fluxes in the $0.2r_{500}-r_{500}$ radial range. 

To select the corresponding flux limit in this radial range, we considered a beta profile with the mean parameters from our analysis and computed the fraction of the flux integrated in this range. From Table \ref{tabpar}, we find $\beta\sim0.64$ with a scatter of  0.13 and $r_{c1}\sim0.16 r_{500}$  with a scatter of $0.08r_{500}$. Integrating a surface-brightness profile with such parameters, we estimate that 59\% of the flux originates from the $0.2r_{500}-r_{500}$, and therefore the HIFLUGCS flux limit translates into a minimum flux of $7.2\times10^{-12}$ ergs cm$^{-2}$ $\mbox{s}^{-1}$ (0.5-2.0 keV band, $0.2r_{500}-r_{500}$). Because of the 15\% normalization difference, this corresponds to a flux limit of $8.3\times 10^{-12}$ ergs cm$^{-2}$ $\mbox{s}^{-1}$ for \textit{XMM-Newton} fluxes.

\subsection{Incompleteness}
\label{incompleteness}

Since our sample is selected as a subset of HIFLUGCS and not directly from RASS data, there may be objects that are not in the HIFLUGCS sample but should be in ours. In other terms, there could be some objects with a flat profile, for which the total flux would be slightly below the HIFLUGCS limit but the flux in the $0.2r_{500}-r_{500}$ radial range would be above our new limit. This effect could have an influence on our measurement of the CC fraction using our subsample, and it must be quantified. 

\begin{figure}
\resizebox{\hsize}{!}{\includegraphics{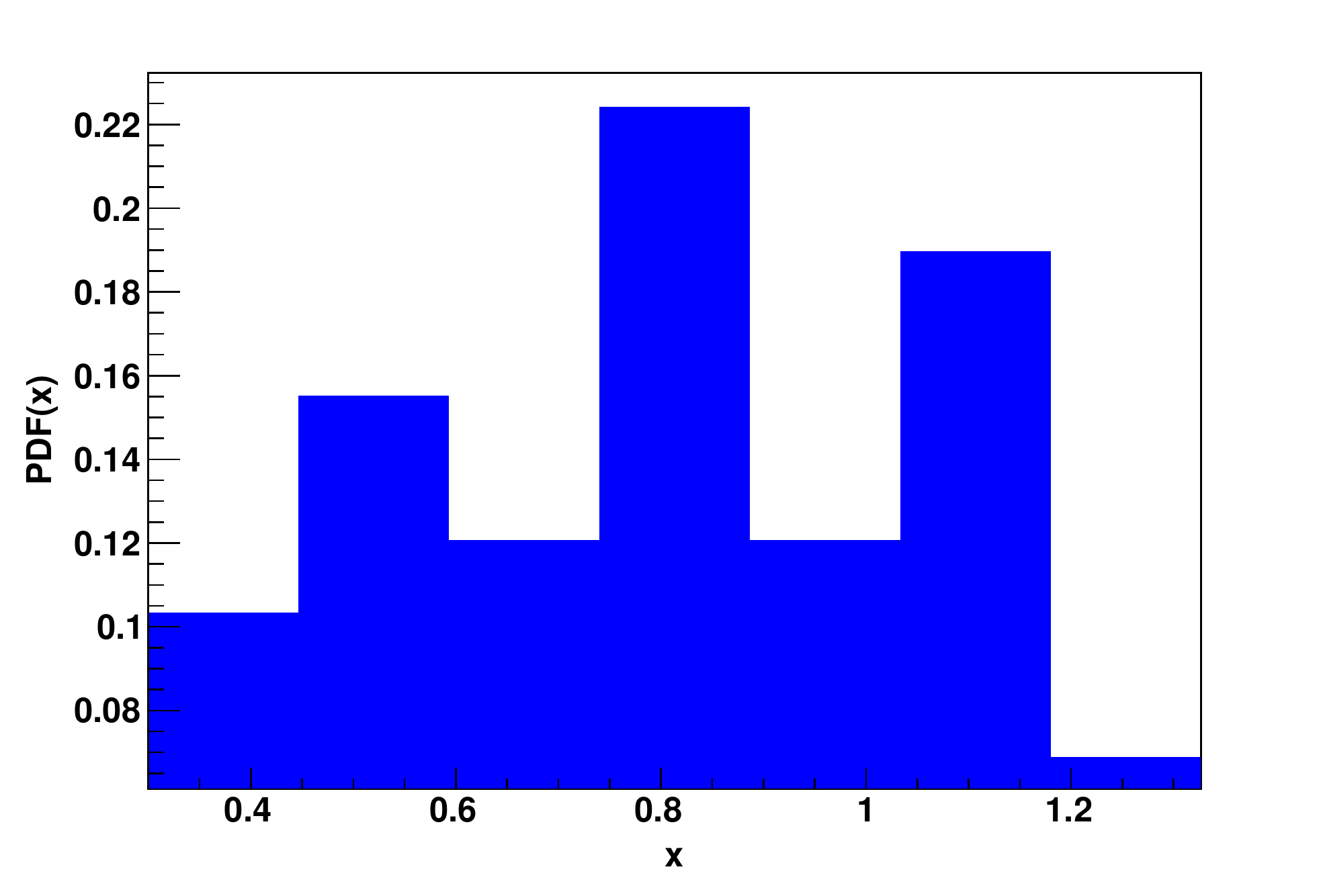}}
\caption{Normalized distribution of the quantity $x=f_{annulus}/f_{old}$, where $f_{annulus}$ is the ratio between the flux computed in the $0.2r_{500}-r_{500}$ radial range and the corresponding flux limit ($7.2\times10^{-12}$ ergs cm$^{-2}$ $s^{-1}$, 0.5-2.0 keV), and $f_{old}$ is the original flux from \citet{reip} rescaled by the corresponding flux limit ($2.0\times10^{-11}$ ergs cm$^{-2}$ $s^{-1}$, 0.1-2.4 keV). This defines a probability distribution $PDF(x)$ for the probability that the flux of a cluster is increased or decreased when performing our new selection.}
\label{pdfofx}
\end{figure}

To estimate the number of objects which might be missed in our subsample, we computed the distribution of the ratio between our fluxes in the $0.2r_{500}-r_{500}$ and the original survey fluxes, excluding the most prominent outliers from Fig. \ref{fnew} and rescaling the distribution by the ratio of the two flux limits. This defines a probability distribution $PDF(x)$ for the probability that the flux of a cluster would be increased or decreased with respect to the flux limit when going from the original selection in the HIFLUGCS sample to the one we apply. In the probability distribution, the peaked objects are the ones with a ratio much below 1, while the flat systems have a ratio larger than 1. In Fig. \ref{pdfofx} we show the resulting probability distribution.

For an object with a given total flux $F<F_{lim}$, the probability to have a $0.2r_{500}-r_{500}$ flux greater than our flux limit is given by

\be P(F)=\int_{F_{lim}/F}^{\infty} PDF(x)\, dx,\ee

\noindent  We can then estimate the number of missed objects by integrating the probability distribution weighted by the number of objects at a given flux, i.e.
\be N_{missed}=\sum_{F<F_{lim}} N_{obj}(F)\times P(F).\ee
\noindent where $N_{obj}$ is estimated from the $\log N-\log S$ of the REFLEX \citep{reflex} and NORASS \citep{norass} samples from which HIFLUGCS is extracted. As a result, the average number of objects which should be selected in our subsample but not in HIFLUGCS is $N_{missed}\sim3.89$. Given that the objects which might be missed are the ones with a high flux outside the core, their surface-brightness profile should be very flat, so we expect them to be mostly NCC. In conclusion, we expect that statistically between 3 and 4 NCC clusters are missed from our selection.

\subsection{Estimate of the CC fraction}
\label{estimate}

After the selection of the clusters based on their flux in the $0.2r_{500}-r_{500}$ radial range, we reject 13 clusters from the original sample. Table \ref{missthecut} shows the corresponding objects and fluxes, along with their classification by H10. Interestingly, all the rejected objects were classified as CC: 9 of the clusters were classified as strong CC clusters (i.e. clusters with a central cooling time $<1$ Gyr, hereafter SCC), and the remaining 4 as weak CC (central cooling time between 1 and 7.7 Gyr, hereafter WCC). Among the rejected WCC clusters, two of them (A2244 and A1650) have a central cooling time just above 1 Gyr, so they are borderline objects which resemble closely strong CC objects. One object, Zw III 54, has a redetermined total flux below the HIFLUGCS cut (see above). As a result, our selection only affected objects which exhibit strong CC characteristics and a prominent central emission excess. We also note that an important fraction of the clusters (5 out of 13, namely EXO 0422, Zw III 54, A 3581, NGC 4636, and Sersic 159-03) have a temperature below 3 keV, i.e. they are low-mass, low-luminosity objects. This is in qualitative agreement with our simulation results (see Fig. \ref{ccfrac}), which show a strong bias of the original sample in the case of low-luminosity objects. Moreover, 2 of the remaining objects (RX J1504 and A2204) are distant ($z>0.15$), strong CC objects, again in agreement with the predictions of our simulations (see the right panel of Fig. \ref{ccfrac}).

\begin{table}
\caption{\label{missthecut}HIFLUGCS clusters rejected for the present work. Column description: 1: Cluster name; 2 and 3: 0.5-2.0 keV fluxes in the radial range $0.2r_{500}-r_{500}$ from \textit{XMM-Newton}/MOS2 (2) and \textit{ROSAT}/PSPC (3), in units of $10^{-12}$ ergs cm$^{-2}$ $\mbox{s}^{-1}$ ; 4: central cooling time in units of $h_{71}^{-1/2}$ Gyr, from H10; 5: cluster classification from H10 (SCC = strong cool-core, WCC = weak cool-core).}
\begin{tabular}{c c c c c}
\hline
\hline
Cluster & $F_{XMM}$ & $F_{ROSAT}$ & CCT & Class\\
\hline
A133  &  6.88$\pm$0.38  &  5.81$\pm$0.24 & 0.47 & SCC\\
A3112  &  7.54$\pm$0.24  &  6.15$\pm$0.21 & 0.37 & SCC\\
Zw III 54  &  4.90$\pm$0.82  &  & 5.48 & WCC\\
EXO 0422  &  5.03$\pm$0.67  &  & 0.47 & SCC\\
NGC 4636  &  2.78$\pm$0.24  &  2.52$\pm$0.22 & 0.21 & SCC\\
A1650  &  5.96$\pm$0.55  &  & 1.25 & WCC\\
A1651  &  8.25$\pm$0.28  &  6.83$\pm$0.20 & 3.63 & WCC\\
A3581  &  5.87$\pm$0.38  &  & 0.52 & SCC\\
RX J1504  &  2.71$\pm$0.19  &  & 0.59 & SCC\\
A2204  &  4.93$\pm$0.38  &  & 0.25 & SCC\\
A2244  &    &  5.02$\pm$0.26 & 1.53 & WCC\\
Sersic 159-03  &  4.61$\pm$0.36  &  3.87$\pm$0.25 & 0.88 & SCC\\
A2597  &  3.64$\pm$0.22  &  2.88$\pm$0.18 & 0.42 & SCC\\
\hline
\end{tabular}

\end{table}

\begin{table}[!ht]
\caption{\label{subsample}Unbiased HIFLUGCS subsample used in the present work. Column description: 1: Cluster name; 2 and 3: 0.5-2.0 keV fluxes in the radial range $0.2r_{500}-r_{500}$ from \textit{XMM-Newton}/MOS2 (2) and \textit{ROSAT}/PSPC (3), in units of $10^{-12}$ ergs cm$^{-2}$ $\mbox{s}^{-1}$; 4: central cooling time in units of $h_{71}^{-1/2}$ Gyr, from H10; 5: cluster classification from H10 (SCC = strong cool-core, WCC = weak cool-core, NCC = non cool-core).}
\begin{tabular}{c c c c c}
\hline
\hline
Cluster & $F_{XMM}$ & $F_{ROSAT}$ & CCT & Class\\
\hline
A85  &  23.45$\pm$0.73  &  20.38$\pm$0.38 & 0.51 & SCC\\
A119  &  22.17$\pm$0.78  &  19.11$\pm$0.43 & 14.03 & NCC\\
NGC 507  &  9.35$\pm$0.36  &  8.32$\pm$0.34 & 0.48 & SCC\\
A262  &  30.52$\pm$1.87  &  24.62$\pm$1.16 & 0.43 & SCC\\
A400  &  11.69$\pm$0.71  &  11.36$\pm$0.31 & 8.04 & NCC\\
A399  &  13.74$\pm$0.85  &  & 12.13 & NCC\\
A401  &  17.99$\pm$0.48  &  16.38$\pm$0.35 & 8.81 & NCC\\
Fornax  &    &  80.88$\pm$10.79 & 0.69 & SCC\\
2A 0335  &  16.03$\pm$1.51  &  12.78$\pm$1.68 & 0.31 & SCC\\
A3158  &  14.18$\pm$0.39  &  12.97$\pm$0.34 & 8.22 & NCC\\
A478  &  10.57$\pm$0.20  &  9.44$\pm$0.14 & 0.43 & SCC\\
NGC 1550  &  24.27$\pm$4.96  &  & 0.23 & SCC\\
A3266  &  26.80$\pm$0.41  &  21.08$\pm$0.36 & 7.62 & WCC\\
A496  &  27.42$\pm$1.37  &  23.92$\pm$1.18 & 0.47 & SCC\\
A3376  &  10.75$\pm$0.68  &  10.71$\pm$0.45 & 16.47 & NCC\\
A3391  &   &  8.667$\pm$0.72 & 12.46 & NCC\\
A3395s  &  11.16$\pm$1.56  &  9.955$\pm$1.23 & 12.66 & NCC\\
A576  &  10.90$\pm$0.81  &  & 3.62 & WCC\\
A754  &  31.25$\pm$0.77  &  24.74$\pm$0.48 & 9.53 & NCC\\
Hydra A &  10.52$\pm$0.21  &  8.96$\pm$0.16 & 0.41 & SCC\\
A1060  &   &  25.81$\pm$4.33 & 2.87 & WCC\\
A1367  &    &  40.50$\pm$1.03 & 27.97 & NCC\\
MKW 4  &   &  6.12$\pm$1.86 & 0.28 & SCC\\
Zw Cl 1215  &  9.09$\pm$0.53  &  & 10.99 & NCC\\
A3526  &   &  72.64$\pm$11.28 & 0.42 & SCC\\
A1644  &  21.77$\pm$1.99  &  & 0.84 & SCC\\
Coma  &    &  128.59$\pm$1.74 & 15.97 & NCC\\
NGC 5044  &   &  12.57$\pm$0.48 & 0.21 & SCC\\
A1736  &  14.65$\pm$0.75  &  & 16.59 & NCC\\
A3558  &  27.38$\pm$0.26  &  26.10$\pm$0.23 & 1.69 & WCC\\
A3562  &  13.81$\pm$0.19  &  11.24$\pm$0.25 & 5.15 & WCC\\
A3571  &  38.61$\pm$2.04  &  34.91$\pm$1.38 & 2.13 & WCC\\
A1795  &  14.97$\pm$0.33  &  12.14$\pm$0.22 & 0.61 & SCC\\
MKW 8  &  9.22$\pm$4.94  &  & 10.87 & NCC\\
A2029  &  16.06$\pm$2.53  &  13.42$\pm$1.57 & 0.53 & SCC\\
A2052  &  11.69$\pm$0.75  &  9.55$\pm$0.61 & 0.51 & SCC\\
MKW 3S  &  8.34$\pm$0.71  &  6.67$\pm$0.53 & 0.86 & SCC\\
A2065  &  9.63$\pm$2.03  &  & 1.34 & WCC\\
A2063  &  12.38$\pm$1.57  &  11.02$\pm$1.18 & 2.36 & WCC\\
A2142  &   &  18.38$\pm$0.33 & 1.94 & WCC\\
A2147  &   &  31.31$\pm$1.26 & 17.04 & NCC\\
A2163  &  8.06$\pm$0.30  &  & 9.65 & NCC\\
A2199  &  26.93$\pm$0.83  &  22.13$\pm$0.49 & 0.60 & SCC\\
A2256  &  26.28$\pm$0.70  &  23.21$\pm$0.36 & 11.56 & NCC\\
A2255  &  9.52$\pm$0.29  &  7.86$\pm$0.14 & 20.66 & NCC\\
A3667  &  35.86$\pm$0.35  &  29.73$\pm$0.27 & 6.14 & WCC\\
A2589  &  8.05$\pm$0.51  &  6.98$\pm$0.35 & 1.18 & WCC\\
A2634  &   &  11.12$\pm$0.86 & 1.52 & WCC\\
A2657  &  9.38$\pm$0.48  &  8.94$\pm$0.21 & 2.68 & WCC\\
A4038  &  18.50$\pm$0.56  &  16.31$\pm$0.45 & 1.68 & WCC\\
A4059  &  10.00$\pm$0.38  &  8.161$\pm$0.32 & 0.7 & SCC\\
\hline
\end{tabular}

\end{table}

After the selection, our subsample comprises 51 objects (see Table \ref{subsample}). The table shows our selection of clusters, together with their fluxes in the  $0.2r_{500}-r_{500}$ annulus, central cooling time and classification from the work of H10. Four clusters (MKW 4, MKW 3S, A2163 and A2589) show a flux below the limit, but consistent with it within $1\sigma$, so we included them in our selection. 

Of the 51 objects in our subsample, 19 (37\%) are classified as strong CC, 18 (35\%) as NCC, and 14 (27\%) are WCC. When taking into account the clusters which might be missed in our subsample (see Sect. \ref{incompleteness}) and assuming that all of them are NCC, the fraction of SCC is reduced to 35\% and that of NCC increased to 40\%. Compared to the analysis of \citet{mittal}, which found 44\% SCC clusters, our value of 35\% corresponds to a bias of 26\% in the original sample. This result is in excellent agreement with the bias estimated from our simulation (29\%, see Sect. \ref{sres}). 

\subsection{The ratio $F_{annulus}/F_{core}$ as a cool-core indicator}
\label{ccind}

\begin{figure}
\resizebox{\hsize}{!}{\includegraphics{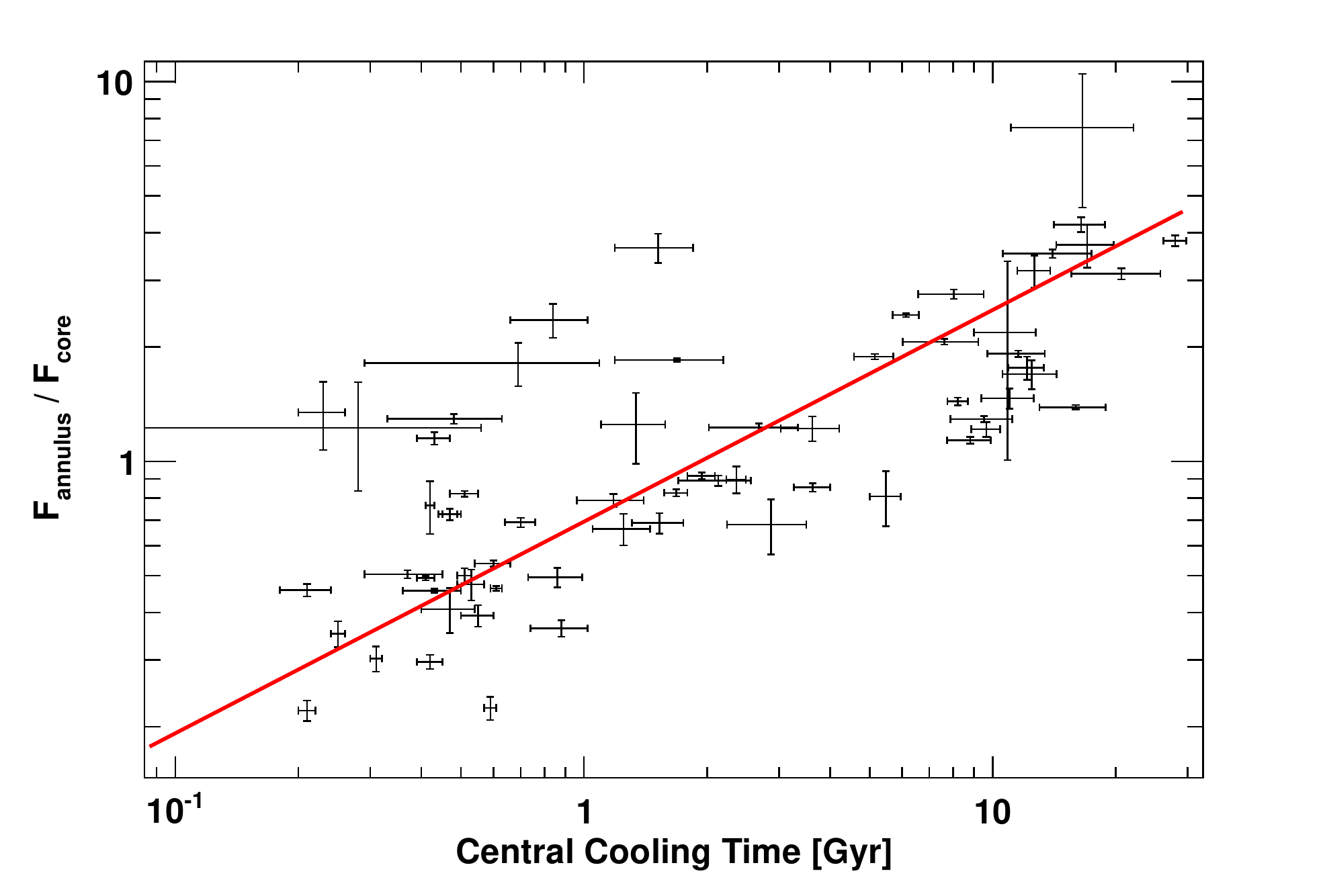}}
\caption{Ratio between the flux in the $0.2r_{500}-r_{500}$ radial range and the flux in the core integrated up to $0.2r_{500}$, $F_{annulus}/F_{core}$ (this work), as a function of the central cooling time in units of Gyr (H10). The red solid line shows a fit to the data with a power law. Spearman's correlation coefficient for the relation is $\rho=0.73^{+0.02}_{-0.05}$.}
\label{fratvscct}
\end{figure}

Given that all the clusters excluded by our analysis present CC characteristics, it is clear that the ratio between the flux in the core, from 0 to $0.2r_{500}$ (hereafter $F_{core}$) and the flux in the annulus $0.2r_{500}-r_{500}$ ($F_{annulus}$) strongly depends on the state of the cluster. In other terms, the ratio $F_{annulus}/F_{core}$ could be used as an indicator of the CC state. To investigate this possibility, we checked for a positive correlation between this quantity and the central cooling time (CCT) as found by H10. In Fig. \ref{fratvscct} we show the ratio $F_{annulus}/F_{core}$ as a function of the CCT for all 64 HIFLUGCS clusters, fitted by a simple power law. A positive correlation is indeed observed, with the best-fit parameters given by

\be \frac{F_{annulus}}{F_{core}}=(0.694\pm0.006)\times \left(\frac{CCT}{1\mbox{ Gyr}}\right)^{0.557\pm0.008}. \label{correlation}\ee

\noindent The correlation between the two quantities is significant: Spearman's correlation factor for the relation is $\rho=0.73^{+0.02}_{-0.05}$. The corresponding likelihood of the correlation occurring by chance is $<10^{-5}$. The scatter of the relation is $\sim24\%$. Given that the CCT is one of the most commonly-used CC indicators, we conclude that the quantity $F_{annulus}/F_{core}$ can indeed be used as an indicator of CC type, and thus Eq. \ref{correlation} can be used to estimate the central cooling time. This indicator is robust, since it does not depend on any deprojected quantity (similar to the ``cuspiness" indicator, \citet{vikhlinin}). However, tight constraints on the surface-brightness profile parameters are necessary, since the fluxes need to be extrapolated to $r_{500}$. This is a problem for very extended, nearby objects, for which the uncertainties when extrapolating to $r_{500}$ are quite large. Indeed, all of the outliers in Fig. \ref{fratvscct} are nearby ($z<0.02$), very extended objects, for which it is difficult to extrapolate the fluxes. Moreover, a number of these objects (NGC 507, A262, NGC 1550, MKW 4) are galaxy groups. Because of the difficulty of observing their outer regions, for groups the measured values of $\beta$ tend to be flatter than the real value \citep{gasta}, which probably results in an over-estimation of $F_{annulus}$. On the other hand, this indicator is excellent for intermediate-redshift objects ($z\sim0.1$), for which the surface-brightness profile parameters can be well constrained, and for which PSF smearing can make deprojection analyses with \emph{XMM-Newton} difficult.

\section{Discussion}

Expanding on the work of H10, our analysis reveals that the flux-limited samples of galaxy clusters, and in particular the HIFLUGCS complete sample, are significantly biased in favor of CC objects, and hence this effect must be taken into account when estimating the fraction of CC objects from a flux-limited sample. Our assessments of the bias through observations and simulations agree very well with each other, which strengthens our point. Overall, we find that $\sim29\%$ of the objects should be removed from the original HIFLUGCS sample if one wishes to estimate the CC fraction, and we extracted a subsample as free as possible of the bias, which can also be used for other purposes (see Table \ref{subsample}). 

Once the appropriate corrections have been applied, we find that considerably less than half of the objects (38\% in our subsample, 35\% when taking incompleteness into account; see Sect. \ref{estimate}) are classified as strong CC clusters, i.e. exhibit a central cooling time below 1 Gyr. This result has important repercussions on the structure formation scenarios. At variance with the original predictions of the cooling-flow model \citep{peres}, in the scenario where CC clusters trace relaxed objects our result implies that a majority of clusters has not reached a stable state. In other terms, in the majority of objects injection of entropy in the ICM by merging events or giant AGN outbursts prevents the formation of cool cores \citep{mcnamara}. 

Obviously, if cool cores cannot be destroyed efficiently by merging events and the state of a cluster is defined once and for all during the cluster formation process, as suggested by some numerical simulations \citep{poole,mccarthy}, the fraction of CC can only be predicted through large cosmological simulations. Conversely, assuming that clusters evolve through time, the CC fraction can be used to estimate roughly the rate of entropy injection events in the local Universe.

Assuming that the NCC fraction evolves through time, the evolution of the fraction of NCC objects can be described as

\be \frac{\mbox{d}f_{NCC}}{\mbox{d}t}=r_{he}f_{CC}-r_{cool}f_{NCC},\label{cevol}\ee

\noindent where $r_{he}$ is the rate of heating events per dark matter halo, $r_{cool}$ is the cooling rate, $f_{NCC}$ is the fraction of both NCC and WCC clusters and $f_{CC}$ is the fraction of SCC clusters. This equation simply reflects the idea that heating events will transform CC clusters into NCC (or WCC), while cooling will do the opposite. The cooling rate is given by $r_{cool}=1/\tau_R$, where the relaxation timescale $\tau_R$ is the time needed for a cluster to relax from a major heating event. In this framework, SCC clusters are the ones which have not experienced any major entropy injection event during the relaxation timescale, while NCC and WCC have experienced or are currently experiencing an entropy injection phase. 

If we assume that the evolution of $f_{NCC}$ is slow, we are close to a stationary situation, and Eq. \ref{cevol} reads
\be r_{he}\sim r_{cool}\frac{f_{NCC}}{f_{CC}}=\frac{1}{\tau_R}\frac{f_{NCC}}{f_{CC}}.\ee
\noindent  From the simulations of \citet{poole2}, it appears that clusters require in average a timescale of $\sim$ 5 Gyr to relax from a merging event. This also corresponds to the mean of the distribution of central cooling times, so we use $\tau_R=5$ Gyr for our calculation. In the $\Lambda$CDM cosmology, a look-back time of 5 Gyr corresponds to a redshift of 0.4, so we are probing the rate of entropy injection events integrated up to redshift 0.4. As a result, we find that the rate of major heating events per cluster is roughly

\be r_{he} \sim 1/3 \mbox{ Gyr}^{-1}. \ee

This number can be compared with the predictions from full cosmological simulations \citep[see e.g.,][]{fakhouri}. Obviously, this estimate relies on several strong assumptions, in particular on the estimate of the relaxation timescale and on the approximation of equilibrium. From a statistical analysis of cold fronts in galaxy clusters, \citet{ghizzardi} estimate a merger rate of $\sim1/3$ Gyr$^{-1}$, similar to ours. These numbers can however not be directly compared, since most of the merging events responsible for producing cold fronts are minor mergers which are not capable of disrupting a cool core. On the other hand, our calculation gives an estimate of the rate of major entropy injection events, whether they are mergers or AGN outbursts, which are capable of transforming a CC cluster into NCC or WCC.

An interesting result of our simulation is the strong dependence of the CC bias on luminosity (see the left panel of Fig. \ref{ccfrac}). While for clusters ($L_X\gtrsim 3\times10^{43}$ ergs $\mbox{s}^{-1}$) the bias is modest in HIFLUGCS, groups and poor clusters appear to be very strongly affected by this effect. Our analysis of the sample qualitatively confirms this result. While only 12 objects (18\%) exhibit a temperature below 3 keV in the complete sample, almost half (5 out of 11) of the objects which were rejected based on our flux cut in the $0.2r_{500}-r_{500}$ annulus fall into this category. As a result, an artificial over-representation of cool cores among low-luminosity objects might be observed when analyzing such a sample. Such a dependence of the CC fraction was indeed observed in HIFLUGCS \citep{chen}. Recently, \citet{johnson} claimed to have identified a population of merging, NCC galaxy groups, which present very shallow surface-brightness profiles, and hence would be difficult to detect in the RASS. This population of NCC groups could have been missed when selecting the sample, thus influencing the observed CC fraction.

A similar, albeit more modest effect impacts on the high-redshift clusters ($z>0.1$). At these redshifts, only the brightest objects are selected, and because of the cut-off in the luminosity function these objects are very rare (see Sect. \ref{sres}). Our selection again qualitatively confirms this result (see Sect. \ref{estimate}). While only three objects in the complete sample have a redshift higher than 0.1, two of them (A2204 and RX J1504) are strong CC objects and are rejected when performing the selection based on the flux in the $0.2 r_{500}-r_{500}$ radial range, the remaining one (A2163) being a very disturbed merging object \citep{a2163}. While this result is not statistically significant given the very small number of objects in this category, it agrees with the predictions of our simulation. This implies that the CC bias should be taken properly into account when studying the cosmological evolution of the CC fraction. 

In addition, we have shown that our fluxes in the $0.2r_{500}-r_{500}$ radial range can be used as a tracer of the state of a cluster when compared to the core flux integrated up to $0.2r_{500}$ (see Sect. \ref{ccind}), since the ratio $F_{annulus}/F_{core}$ correlates with the central cooling time. This relation can be interpreted as the fact that, while the profiles are very different in the cores, they are essentially self-similar in the outer regions of clusters \citep{neumann,lrm09}. Indeed, while in the central regions non-gravitational effects (radiative cooling, AGN feedback, ...) are important, in cluster outskirts the profiles are mostly determined by gravitational processes. Highlighting the importance of excluding the cores when analyzing gravitational effects, \citet{maughan} observed that the $M-T$ relation tightens when these quantities are measured beyond $0.15r_{500}$. The ratio $F_{annulus}/F_{core}$ therefore traces the deviations from the expectations of the self-similar model, and thus can be used to trace the state of a cluster. This suggests that for the extraction of quantities depending on gravitational processes only, such as the $M-T$ and $L_X-T$ relations, the use of a sample selected using fluxes in a well-defined radial range excluding the cores is important \citep[see e.g.,][]{ota}.

\section{Conclusion}

In this paper, we presented an analysis of the effect of the different surface-brightness profiles for CC and NCC clusters on the selection of X-ray flux-limited samples, based on both numerical and observational approaches, and estimated the fraction of CC objects in the nearby Universe. Our results can be summarized as follows:

\begin{itemize}
\item
Performing realistic simulations of a population of clusters and of the selection process of the HIFLUGCS sample, we estimate that $\sim29$\% of the strong CC objects present in the sample should be removed if one wishes to measure the fraction of CC vs NCC objects using this sample.

\item
Analyzing the populations of simulated CC and NCC clusters in the $L_X-z$ plane, we see that CC clusters populate a broader range in the diagram. In other terms, the flux limit is actually different for CC and NCC objects when selecting the clusters based on their observed flux only.

\item
In all cases, we find that low-luminosity objects (groups, poor clusters) are much more affected than more luminous objects. This effect might explain the lack of NCC objects in group samples extracted from RASS data noted by several authors \citep[e.g.,][]{chen}. This illustrates the importance of taking into account this bias when computing the CC fraction. We also find a trend of increasing bias with redshift.

\item
From our analysis of the surface-brightness profiles of all HIFLUGCS clusters, we propose to select clusters according to their flux in a well-defined physical radial range excluding the core ($0.2r_{500}-r_{500}$). Performing a new selection according to the flux in this radial range, we exclude 13 objects from the original sample, all of which present CC characteristics, and extract a subsample of HIFLUGCS as free as possible of the CC bias.

\item Less than half ($35-37\%$) of the clusters in our subsample exhibit strong CC properties. Relating the observed CC fraction with the rate of major entropy injection events in the local Universe, we give a rough estimate of $\sim0.3$ Gyr$^{-1}$ for the rate of heating events per dark-matter halo and per Gyr.

\item
In addition, we find that the ratio $F_{annulus}/F_{core}$ between the fluxes in the radial range $0.2r_{500}-r_{500}$ and $0-0.2r_{500}$ strongly correlates with the central cooling time, and therefore can be used as a CC indicator. We expect this indicator to be particularly effective for intermediate-redshift objects ($z\sim0.1$).

\end{itemize}

\begin{table*}
\caption{\label{tabobs}Basic properties of HIFLUGCS clusters. Column description: 1: Cluster name; 2: instrument used (1=\textit{ROSAT}; 2=\textit{XMM-Newton}; 3=combination of the two); 3: redshift, from \citet{reip} and references therein; 4: virial temperature in keV, from H10; 5 and 6: $r_{500}$ from the scaling relations of \citet{arnaud} in kpc (5), and the subtended angle $\theta_{500}$ in arcmin (6).}
\begin{center}
\begin{tabular}{c c c c c c }
\hline\hline
Cluster & Instrument & Redshift & $kT_{vir}$ [kev] & $r_{500}$ [kpc] & $\theta_{500}$ [arcmin] \\
\hline
A85  &  3  &  0.0556  &  6  &  1208  &  19.2\\
A119  &  3  &  0.044  &  5.73  &  1187  &  23.5\\
A133  &  3  &  0.0569  &  3.96  &  981  &  15.2\\
NGC 507  &  3  &  0.0165  &  1.44  &  539  &  27.5\\
A262  &  3  &  0.0161  &  2.44  &  729  &  38.1\\
A400  &  3  &  0.024  &  2.26  &  695  &  24.6\\
A399  &  2  &  0.0715  &  6.7  &  1268  &  15.9\\
A401  &  3  &  0.0748  &  8.51  &  1427  &  17.2\\
A3112  &  3  &  0.075  &  4.73  &  1064  &  12.8\\
Fornax  &  1  &  0.0046  &  1.34  &  520  &  93.9\\
2A 0335  &  3  &  0.0349  &  3.53  &  935  &  23.1\\
Zw III 54  &  2  &  0.0311  &  2.5  &  734  &  20.2\\
A3158  &  3  &  0.059  &  4.99  &  1100  &  16.5\\
A478  &  3  &  0.09  &  7.34  &  1316  &  13.4\\
NGC 1550  &  2  &  0.0123  &  1.34  &  519  &  35.3\\
EXO 0422  &  2  &  0.039  &  2.93  &  801  &  17.8\\
A3266  &  3  &  0.0594  &  9.45  &  1514  &  22.6\\
A496  &  3  &  0.0328  &  4.86  &  1098  &  28.7\\
A3376  &  3  &  0.0455  &  3.8  &  966  &  18.5\\
A3391  &  1  &  0.0531  &  5.77  &  1186  &  19.6\\
A3395S  &  3  &  0.0498  &  4.82  &  1086  &  19.1\\
A576  &  2  &  0.0381  &  4.09  &  1005  &  22.8\\
A754  &  3  &  0.0528  &  11.13  &  1648  &  27.4\\
Hydra A  &  3  &  0.0538  &  3.45  &  874  &  14.3\\
A1060  &  1  &  0.0114  &  3.16  &  846  &  62.1\\
A1367  &  1  &  0.0216  &  3.58  &  947  &  37.1\\
MKW 4  &  1  &  0.02  &  2.01  &  651  &  27.5\\
Zw Cl 1215  &  2  &  0.075  &  6.27  &  1225  &  14.7\\
NGC 4636  &  3  &  0.0037  &  0.9  &  415  &  93.0\\
A3526  &  1  &  0.0103  &  3.92  &  996  &  80.8\\
A1644  &  2  &  0.0474  &  5.09  &  1117  &  20.6\\
A1650  &  2  &  0.0845  &  5.81  &  1174  &  12.7\\
A1651  &  3  &  0.086  &  6.34  &  1226  &  13.0\\
Coma  &  1  &  0.0232  &  9.15  &  1513  &  55.3\\
NGC 5044  &  1  &  0.009  &  1.22  &  492  &  45.6\\
A1736  &  2  &  0.0461  &  3.12  &  828  &  15.7\\
A3558  &  3  &  0.048  &  4.95  &  1101  &  20.0\\
A3562  &  3  &  0.0499  &  4.43  &  1041  &  18.3\\
A3571  &  3  &  0.0397  &  7  &  1314  &  28.6\\
A1795  &  3  &  0.0616  &  6.08  &  1213  &  17.5\\
A3581  &  2  &  0.0214  &  1.97  &  644  &  25.5\\
MKW 8  &  2  &  0.027  &  3  &  816  &  26.0\\
RX J1504  &  2  &  0.2153  &  9.53  &  1414  &  6.9\\
A2029  &  3  &  0.0767  &  8.26  &  1405  &  16.5\\
A2052  &  3  &  0.0348  &  3.35  &  866  &  21.4\\
MKW 3  &  3  &  0.045  &  3.9  &  979  &  18.9\\
A2065  &  2  &  0.0721  &  5.4  &  1138  &  14.2\\
A2063  &  3  &  0.0354  &  3.77  &  966  &  23.5\\
A2142  &  1  &  0.0899  &  8.4  &  1408  &  14.4\\
A2147  &  1  &  0.0351  &  4.07  &  1004  &  24.6\\
A2163  &  2  &  0.201  &  15.91  &  1840  &  9.5\\
A2199  &  3  &  0.0302  &  4.37  &  1042  &  29.5\\
A2204  &  2  &  0.152  &  8.92  &  1411  &  9.1\\
A2244  &  1  &  0.097  &  5.78  &  1165  &  11.1\\
A2256  &  3  &  0.0601  &  7.61  &  1358  &  20.0\\
A2255  &  3  &  0.08  &  5.81  &  1176  &  13.3\\
A3667  &  3  &  0.056  &  6.39  &  1247  &  19.6\\
Sersic 159-03  &  3  &  0.058  &  2.57  &  737  &  11.2\\
A2589  &  3  &  0.0416  &  3.89  &  979  &  20.4\\
A2597  &  3  &  0.0852  &  4.05  &  980  &  10.5\\
A2634  &  1  &  0.0312  &  3.19  &  844  &  23.2\\
A2657  &  3  &  0.0404  &  3.52  &  932  &  20.0\\
A4038  &  3  &  0.0283  &  3.14  &  837  &  25.3\\
A4059  &  3  &  0.046  &  4.22  &  1018  &  19.3\\
\hline
\end{tabular}

\end{center}
\end{table*}

\begin{table*}
\caption{\label{tabpar}Results of surface-brightness profile fitting. Column description: 1: Cluster name; 2: $\beta$; 3: outer core radius $r_{c1}$ (in kpc); 4: inner core radius $r_{c2}$ (in kpc); 5: ratio between the two beta components at $r=0$, $R$; 6 and 7: estimated total fluxes from 0 to $r_{500}$ in the 0.5-2.0 keV band, in units of $10^{-11}$ ergs cm$^{-2}$ $\mbox{s}^{-1}$, from \textit{XMM-Newton}/MOS2 (6) and \textit{ROSAT}/PSPC (7); NB: a cross-calibration factor of 0.85 must be applied to the \textit{XMM-Newton} fluxes (see Appendix \ref{calib}) ; 8: minimum $\chi^2$ of the fit versus number of degrees of freedom.}
\begin{center}
\begin{tabular}{c c c c c c c c}
\hline
\hline
Cluster & $\beta$ & $r_{c1}$ & $r_{c2}$ & $R$ & $F_{XMM}$ & $F_{ROSAT}$ & $\chi^2$/d.o.f. \\
\hline
A85  &  0.638$\pm$0.009  &  192$\pm$8  &  35$\pm$2  &  11.2$\pm$0.5  &  5.175$\pm$0.072  &  4.540$\pm$0.038  &  349.2/222\\
A119  &  0.647$\pm$0.015  &  363$\pm$11  &    &    &  2.834$\pm$0.076  &  2.451$\pm$0.042  &  309.8/229\\
A133  &  0.579$\pm$0.009  &  145$\pm$11  &  30$\pm$1  &  23.8$\pm$2.7  &  1.626$\pm$0.038  &  1.392$\pm$0.023  &  302.9/220\\
NGC 507  &  0.458$\pm$0.003  &  16.4$\pm$0.5  &    &    &  1.625$\pm$0.036  &  1.461$\pm$0.034  &  784.5/261\\
A262  &  0.531$\pm$0.008  &  85$\pm$3  &  8.0$\pm$0.3  &  22.6$\pm$0.9  &  5.622$\pm$0.187  &  4.522$\pm$0.115  &  428.6/229\\
A400  &  0.491$\pm$0.008  &  94$\pm$4  &    &    &  1.616$\pm$0.070  &  1.568$\pm$0.031  &  339.4/229\\
A399  &  0.599$\pm$0.031  &  210$\pm$14  &    &    &  2.111$\pm$0.085  &    &  101.7/96\\
A401  &  0.679$\pm$0.009  &  222$\pm$5  &    &    &  3.372$\pm$0.048  &  3.098$\pm$0.034  &  313.8/212\\
A3112  &  0.608$\pm$0.006  &  88$\pm$4  &  23$\pm$1  &  6.6$\pm$0.6  &  2.222$\pm$0.023  &  1.855$\pm$0.021  &  263.9/219\\
Fornax  &  0.685$\pm$0.033  &  135$\pm$4  &  12.3$\pm$0.9  &  8.4$\pm$0.4  &    &  11.980$\pm$1.062  &  501.9/142\\
2A 0335  &  0.659$\pm$0.015  &  125$\pm$7  &  26.0$\pm$0.8  &  52$\pm$7.2  &  7.000$\pm$0.150  &  5.655$\pm$0.166  &  285.4/203\\
Zw III 54  &  0.703$\pm$0.029  &  99$\pm$5  &    &    &  1.092$\pm$0.081  &    &  115.1/115\\
A3158  &  0.632$\pm$0.01  &  172$\pm$4  &    &    &  2.41$\pm$0.039  &  2.196$\pm$0.033  &  313/193\\
A478  &  0.662$\pm$0.004  &  121$\pm$2  &  31.1$\pm$0.9  &  3.7$\pm$0.1  &  3.394$\pm$0.020  &  3.028$\pm$0.013  &  592.4/229\\
NGC 1550  &  0.436$\pm$0.012  &  21$\pm$4  &  2.8$\pm$0.2  &  36.1$\pm$3.1  &  4.206$\pm$0.501  &    &  127.8/114\\
EXO 0422  &  0.635$\pm$0.016  &  71$\pm$6  &  21$\pm$1  &  7.5$\pm$0.9  &  1.734$\pm$0.067  &    &  148.7/114\\
A3266  &  0.783$\pm$0.013  &  413$\pm$9  &    &    &  4.061$\pm$0.040  &  3.298$\pm$0.036  &  795.5/246\\
A496  &  0.544$\pm$0.006  &  96$\pm$5  &  18$\pm$0.2  &  17.9$\pm$0.9  &  6.545$\pm$0.138  &  5.614$\pm$0.115  &  200.8/144\\
A3376  &  1.26$\pm$0.077  &  625$\pm$31  &    &    &  1.344$\pm$0.069  &  1.339$\pm$0.044  &  137.9/115\\
A3391  &  0.566$\pm$0.015  &  163$\pm$10  &    &    &    &  1.371$\pm$0.071  &  201.8/146\\
A3395S  &  0.484$\pm$0.023  &  173$\pm$14  &    &    &  1.465$\pm$0.155  &  1.311$\pm$0.124  &  273.2/172\\
A576  &  0.57$\pm$0.013  &  107$\pm$6  &    &    &  1.911$\pm$0.079  &    &  179.7/108\\
A754  &  0.848$\pm$0.015  &  406$\pm$8  &    &    &  5.488$\pm$0.078  &  4.368$\pm$0.048  &  259.8/116\\
Hydra A  &  0.802$\pm$0.009  &  173$\pm$3  &  38.9$\pm$0.6  &  11.8$\pm$0.2  &  3.177$\pm$0.021  &  2.741$\pm$0.015  &  607.9/211\\
A1060  &  0.667$\pm$0.021  &  101$\pm$7  &  27$\pm$3  &  1.8$\pm$0.2  &    &  6.391$\pm$0.431  &  163.4/144\\
A1367  &  0.619$\pm$0.017  &  290$\pm$10  &    &    &    &  5.077$\pm$0.104  &  353.1/146\\
MKW 4  &  0.76$\pm$0.076  &  186$\pm$25  &  41$\pm$6  &  6.2$\pm$1  &    &  1.101$\pm$0.192  &  130.2/112\\
Zw Cl 1215  &  0.652$\pm$0.021  &  209$\pm$9  &    &    &  1.522$\pm$0.053  &    &  136.8/116\\
NGC 4636  &  0.548$\pm$0.003  &  3.1$\pm$0.1  &  0.7$\pm$0.04  &  2.6$\pm$0.2  &  1.571$\pm$0.045  &  1.361$\pm$0.043  &  1951./259\\
A3526  &  0.633$\pm$0.02  &  149$\pm$5  &  20.0$\pm$0.9  &  16.9$\pm$0.8  &   &  16.16$\pm$1.145  &  275.4/144\\
A1644  &  0.448$\pm$0.031  &  122$\pm$23  &  17$\pm$4  &  3.6$\pm$1.5  &  3.043$\pm$0.197  &    &  141.8/107\\
A1650  &  0.866$\pm$0.045  &  276$\pm$20  &  83$\pm$5  &  4.5$\pm$0.3  &  1.494$\pm$0.056  &    &  157.4/114\\
A1651  &  0.637$\pm$0.008  &  134$\pm$4  &    &    &  1.793$\pm$0.027  &  1.487$\pm$0.020  &  219.3/159\\
Coma  &  0.651$\pm$0.005  &  249$\pm$3  &    &    &    &  22$\pm$0.173  &  397/146\\
NGC 5044  &  0.529$\pm$0.002  &  8.5$\pm$0.1  &    &    &    &  4.000$\pm$0.048  &  657.1/146\\
A1736  &  0.67$^4$  &  360$\pm$4  &    &    &  1.746$\pm$0.074  &    &  223.6/107\\
A3558  &  0.551$\pm$0.003  &  152$\pm$2  &    &    &  4.198$\pm$0.025  &  4.020$\pm$0.023  &  637.5/216\\
A3562  &  0.461$\pm$0.003  &  69$\pm$2  &    &    &  2.12$\pm$0.019  &  1.734$\pm$0.025  &  557.5/248\\
A3571  &  0.639$\pm$0.01  &  167$\pm$8  &  58$\pm$5  &  1.2$\pm$0.2  &  8.227$\pm$0.206  &  7.417$\pm$0.139  &  259.5/229\\
A1795  &  0.724$\pm$0.008  &  210$\pm$5  &  54$\pm$1  &  14$\pm$0.5  &  4.718$\pm$0.033  &  3.853$\pm$0.022  &  304.7/199\\
A3581  &  0.569$\pm$0.004  &  18.8$\pm$0.4  &    &    &  2.083$\pm$0.038  &    &  242.5/116\\
MKW 8  &  0.498$\pm$0.051  &  94$\pm$13  &    &    &  1.362$\pm$0.490  &    &  128.3/113\\
RX J1504  &  0.759$\pm$0.018  &  161$\pm$12  &  53$\pm$2  &  11.4$\pm$1.6  &  1.459$\pm$0.018  &    &  101.4/79\\
A2029  &  0.616$\pm$0.009  &  111$\pm$17  &  43$\pm$6  &  3.6$\pm$1.8  &  4.967$\pm$0.254  &  4.159$\pm$0.158  &  195.3/210\\
A2052  &  0.749$\pm$0.017  &  159$\pm$4  &  31.9$\pm$0.8  &  15.7$\pm$0.5  &  3.461$\pm$0.074  &  2.847$\pm$0.060  &  313.8/229\\
MKW 3  &  0.628$\pm$0.009  &  86$\pm$5  &  27$\pm$2  &  3.6$\pm$0.4  &  2.486$\pm$0.070  &  2.014$\pm$0.053  &  338.1/229\\
A2065  &  0.755$\pm$0.086  &  327$\pm$50 &  93$\pm$13  &  4.4$\pm$0.5  &  1.716$\pm$0.212  &    &  155.2/111\\
A2063  &  0.734$\pm$0.027  &  194$\pm$13  &  54$\pm$4  &  3.6$\pm$0.3  &  2.597$\pm$0.155  &  2.330$\pm$0.116  &  266.2/223\\
A2142  &  0.757$\pm$0.021  &  425$\pm$28 &  140$\pm$5  &  9$\pm$0.9  &    &  3.821$\pm$0.033  &  148.3/99\\
A2147  &  0.369$\pm$0.01  &  61$\pm$5  &    &    &    &  3.921$\pm$0.125  &  307.4/190\\
A2163  &  0.628$\pm$0.016  &  253$\pm$12  &    &   &  1.465$\pm$0.029  &    &  89.3/72\\
A2199  &  0.625$\pm$0.004  &  99$\pm$3  &  31$\pm$1  &  4.6$\pm$0.3  &  7.696$\pm$0.084  &  6.376$\pm$0.049  &  393.1/229\\
A2204  &  0.663$\pm$0.013  &  134$\pm$9  &  31$\pm$1  &  12.7$\pm$1.3  &  1.874$\pm$0.037  &    &  145.3/114\\
A2244  &  0.627$\pm$0.011  &  103$\pm$5  &    &    &    &  1.229$\pm$0.026  &  106.5/86\\
A2256  &  0.814$\pm$0.015  &  391$\pm$9  &    &    &  3.954$\pm$0.069  &  3.498$\pm$0.036  &  527.5/231\\
A2255  &  0.795$\pm$0.018  &  424$\pm$13  &    &    &  1.264$\pm$0.029  &  1.054$\pm$0.014  &  300.6/207\\
A3667  &  0.529$\pm$0.004  &  199$\pm$3  &    &    &  5.024$\pm$0.035  &  4.157$\pm$0.027  &  936.3/261\\
Sersic 159-03  &  0.733$\pm$0.02  &  118$\pm$13  &  40$\pm$2  &  13.7$\pm$2.8  &  1.718$\pm$0.035  &  1.444$\pm$0.025  &  433.7/228\\
A2589  &  0.625$\pm$0.01  &  102$\pm$4  &  21$\pm$3  &  1.4$\pm$0.1  &  1.824$\pm$0.051  &  1.583$\pm$0.035  &  246.1/229\\
A2597  &  0.708$\pm$0.016  &  138$\pm$11  &  41$\pm$1  &  20.9$\pm$2.6  &  1.545$\pm$0.022  &  1.256$\pm$0.017  &  315.8/184\\
A2634  &  0.626$\pm$0.037  &  259$\pm$21  &    &    &    &  1.398$\pm$0.086  &  158.2/116\\
A2657  &  0.539$\pm$0.005  &  77$\pm$2  &    &    &  1.717$\pm$0.047  &  1.611$\pm$0.020  &  461.3/230\\
A4038  &  0.53$\pm$0.004  &  39.7$\pm$0.7  &    &    &  4.085$\pm$0.056  &  3.591$\pm$0.045  &  286.7/216\\
A4059  &  0.577$\pm$0.006  &  71$\pm$2  &  13$\pm$1  &  2.2$\pm$0.1  &  2.416$\pm$0.038  &  1.965$\pm$0.031  &  259.5/224\\
\hline
\end{tabular}
\end{center}
\hspace{1cm}$^4$See Appendix \ref{indiv}.
\end{table*}

\acknowledgements{We thank Thomas Reiprich and Fabio Gastaldello for their useful comments. DE is supported by the Occhialini post-doc fellowship grant of IASF Milano. For the present work, we made extensive use of the \textit{XMM-Newton} Science Archive (XSA) and of the High-Energy Astrophysics Science Archive (HEASARC). }

\bibliographystyle{aa}
\bibliography{aa15856}

\begin{thebibliography}{53}
\expandafter\ifx\csname natexlab\endcsname\relax\def\natexlab#1{#1}\fi

\bibitem[{{Andersson} {et~al.}(2009){Andersson}, {Peterson}, {Madejski}, \&
  {Goobar}}]{andersson}
{Andersson}, K., {Peterson}, J.~R., {Madejski}, G., \& {Goobar}, A. 2009, \apj,
  696, 1029

\bibitem[{{Arnaud}(1996)}]{xspec}
{Arnaud}, K.~A. 1996, in Astronomical Society of the Pacific Conference Series,
  Vol. 101, Astronomical Data Analysis Software and Systems V, ed. G.~H.
  {Jacoby} \& J.~{Barnes}, 17--+

\bibitem[{{Arnaud} {et~al.}(2005){Arnaud}, {Pointecouteau}, \&
  {Pratt}}]{arnaud}
{Arnaud}, M., {Pointecouteau}, E., \& {Pratt}, G.~W. 2005, \aap, 441, 893

\bibitem[{{Bagchi} {et~al.}(2006){Bagchi}, {Durret}, {Neto}, \&
  {Paul}}]{bagchi}
{Bagchi}, J., {Durret}, F., {Neto}, G.~B.~L., \& {Paul}, S. 2006, Science, 314,
  791

\bibitem[{{B{\"o}hringer} {et~al.}(2004){B{\"o}hringer}, {Schuecker}, {Guzzo},
  {Collins}, {Voges}, {Cruddace}, {Ortiz-Gil}, {Chincarini}, {De Grandi},
  {Edge}, {MacGillivray}, {Neumann}, {Schindler}, \& {Shaver}}]{reflex}
{B{\"o}hringer}, H., {Schuecker}, P., {Guzzo}, L., {et~al.} 2004, \aap, 425,
  367

\bibitem[{{B{\"o}hringer} {et~al.}(2000){B{\"o}hringer}, {Voges}, {Huchra},
  {McLean}, {Giacconi}, {Rosati}, {Burg}, {Mader}, {Schuecker}, {Simi{\c c}},
  {Komossa}, {Reiprich}, {Retzlaff}, \& {Tr{\"u}mper}}]{norass}
{B{\"o}hringer}, H., {Voges}, W., {Huchra}, J.~P., {et~al.} 2000, \apjs, 129,
  435

\bibitem[{{Cavagnolo} {et~al.}(2009){Cavagnolo}, {Donahue}, {Voit}, \&
  {Sun}}]{cavagnolo}
{Cavagnolo}, K.~W., {Donahue}, M., {Voit}, G.~M., \& {Sun}, M. 2009, \apjs,
  182, 12

\bibitem[{{Cavaliere} \& {Fusco-Femiano}(1976)}]{cavaliere}
{Cavaliere}, A. \& {Fusco-Femiano}, R. 1976, \aap, 49, 137

\bibitem[{{Chen} {et~al.}(2007){Chen}, {Reiprich}, {B{\"o}hringer}, {Ikebe}, \&
  {Zhang}}]{chen}
{Chen}, Y., {Reiprich}, T.~H., {B{\"o}hringer}, H., {Ikebe}, Y., \& {Zhang}, Y.
  2007, \aap, 466, 805

\bibitem[{{Croston} {et~al.}(2008){Croston}, {Pratt}, {B{\"o}hringer},
  {Arnaud}, {Pointecouteau}, {Ponman}, {Sanderson}, {Temple}, {Bower}, \&
  {Donahue}}]{croston}
{Croston}, J.~H., {Pratt}, G.~W., {B{\"o}hringer}, H., {et~al.} 2008, \aap,
  487, 431

\bibitem[{{Cruddace} {et~al.}(2002){Cruddace}, {Voges}, {B{\"o}hringer},
  {Collins}, {Romer}, {MacGillivray}, {Yentis}, {Schuecker}, {Ebeling}, \& {De
  Grandi}}]{cruddace}
{Cruddace}, R., {Voges}, W., {B{\"o}hringer}, H., {et~al.} 2002, \apjs, 140,
  239

\bibitem[{{Donnelly} {et~al.}(2001){Donnelly}, {Forman}, {Jones}, {Quintana},
  {Ramirez}, {Churazov}, \& {Gilfanov}}]{donnelly}
{Donnelly}, R.~H., {Forman}, W., {Jones}, C., {et~al.} 2001, \apj, 562, 254

\bibitem[{{Ebeling} {et~al.}(1997){Ebeling}, {Edge}, {Fabian}, {Allen},
  {Crawford}, \& {Boehringer}}]{ebeling}
{Ebeling}, H., {Edge}, A.~C., {Fabian}, A.~C., {et~al.} 1997, \apjl, 479, L101+

\bibitem[{{Ettori} {et~al.}(2010){Ettori}, {Gastaldello}, {Leccardi},
  {Molendi}, {Rossetti}, {Buote}, \& {Meneghetti}}]{ettori10}
{Ettori}, S., {Gastaldello}, F., {Leccardi}, A., {et~al.} 2010, ArXiv e-prints

\bibitem[{{Fabian}(1994)}]{fabian}
{Fabian}, A.~C. 1994, \araa, 32, 277

\bibitem[{{Fakhouri} \& {Ma}(2008)}]{fakhouri}
{Fakhouri}, O. \& {Ma}, C. 2008, \mnras, 386, 577

\bibitem[{{Gastaldello} {et~al.}(2007){Gastaldello}, {Buote}, {Humphrey},
  {Zappacosta}, {Bullock}, {Brighenti}, \& {Mathews}}]{gasta}
{Gastaldello}, F., {Buote}, D.~A., {Humphrey}, P.~J., {et~al.} 2007, \apj, 669,
  158

\bibitem[{{Ghizzardi} {et~al.}(2010){Ghizzardi}, {Rossetti}, \&
  {Molendi}}]{ghizzardi}
{Ghizzardi}, S., {Rossetti}, M., \& {Molendi}, S. 2010, \aap, 516, A32+

\bibitem[{{Giacconi} {et~al.}(2009){Giacconi}, {Borgani}, {Rosati}, {Tozzi},
  {Gilli}, {Murray}, {Paolillo}, {Pareschi}, {Tagliaferri}, {Ptak},
  {Vikhlinin}, {Flanagan}, {Weisskopf}, {Bignamini}, {Donahue}, {Evrard},
  {Forman}, {Jones}, {Molendi}, {Santos}, \& {Voit}}]{wfxt}
{Giacconi}, R., {Borgani}, S., {Rosati}, P., {et~al.} 2009, in Astronomy, Vol.
  2010, astro2010: The Astronomy and Astrophysics Decadal Survey, 90--+

\bibitem[{{Helsdon} \& {Ponman}(2000)}]{helsdon}
{Helsdon}, S.~F. \& {Ponman}, T.~J. 2000, \mnras, 315, 356

\bibitem[{{Henry} {et~al.}(2004){Henry}, {Finoguenov}, \& {Briel}}]{henry}
{Henry}, J.~P., {Finoguenov}, A., \& {Briel}, U.~G. 2004, \apj, 615, 181

\bibitem[{{Hudson} {et~al.}(2010){Hudson}, {Mittal}, {Reiprich}, {Nulsen},
  {Andernach}, \& {Sarazin}}]{hudson}
{Hudson}, D.~S., {Mittal}, R., {Reiprich}, T.~H., {et~al.} 2010, \aap, 513,
  A37+

\bibitem[{{Johnson} {et~al.}(2009){Johnson}, {Ponman}, \&
  {Finoguenov}}]{johnson}
{Johnson}, R., {Ponman}, T.~J., \& {Finoguenov}, A. 2009, \mnras, 395, 1287

\bibitem[{{Kaastra} {et~al.}(2001){Kaastra}, {Ferrigno}, {Tamura}, {Paerels},
  {Peterson}, \& {Mittaz}}]{kaastra01}
{Kaastra}, J.~S., {Ferrigno}, C., {Tamura}, T., {et~al.} 2001, \aap, 365, L99

\bibitem[{{Kaastra} \& {Mewe}(2000)}]{kaastra}
{Kaastra}, J.~S. \& {Mewe}, R. 2000, in Atomic Data Needs for X-ray Astronomy,
  p. 161, ed. M.~A. {Bautista}, T.~R. {Kallman}, \& A.~K. {Pradhan}, 161--+

\bibitem[{{Kalberla} {et~al.}(2005){Kalberla}, {Burton}, {Hartmann}, {Arnal},
  {Bajaja}, {Morras}, \& {P{\"o}ppel}}]{kalberla}
{Kalberla}, P.~M.~W., {Burton}, W.~B., {Hartmann}, D., {et~al.} 2005, \aap,
  440, 775

\bibitem[{{Kempner} {et~al.}(2002){Kempner}, {Sarazin}, \& {Ricker}}]{kempner}
{Kempner}, J.~C., {Sarazin}, C.~L., \& {Ricker}, P.~M. 2002, \apj, 579, 236

\bibitem[{{Leccardi} \& {Molendi}(2008)}]{lm08}
{Leccardi}, A. \& {Molendi}, S. 2008, \aap, 486, 359

\bibitem[{{Leccardi} {et~al.}(2010){Leccardi}, {Rossetti}, \&
  {Molendi}}]{lrm09}
{Leccardi}, A., {Rossetti}, M., \& {Molendi}, S. 2010, \aap, 510, A82+

\bibitem[{{Markevitch} \& {Vikhlinin}(2001)}]{a2163}
{Markevitch}, M. \& {Vikhlinin}, A. 2001, \apj, 563, 95

\bibitem[{{Maughan}(2007)}]{maughan}
{Maughan}, B.~J. 2007, \apj, 668, 772

\bibitem[{{McCarthy} {et~al.}(2008){McCarthy}, {Babul}, {Bower}, \&
  {Balogh}}]{mccarthy}
{McCarthy}, I.~G., {Babul}, A., {Bower}, R.~G., \& {Balogh}, M.~L. 2008,
  \mnras, 386, 1309

\bibitem[{{McCarthy} {et~al.}(2004){McCarthy}, {Balogh}, {Babul}, {Poole}, \&
  {Horner}}]{mccarthy04}
{McCarthy}, I.~G., {Balogh}, M.~L., {Babul}, A., {Poole}, G.~B., \& {Horner},
  D.~J. 2004, \apj, 613, 811

\bibitem[{{McNamara} \& {Nulsen}(2007)}]{mcnamara}
{McNamara}, B.~R. \& {Nulsen}, P.~E.~J. 2007, \araa, 45, 117

\bibitem[{{Mittal} {et~al.}(2009){Mittal}, {Hudson}, {Reiprich}, \&
  {Clarke}}]{mittal}
{Mittal}, R., {Hudson}, D.~S., {Reiprich}, T.~H., \& {Clarke}, T. 2009, \aap,
  501, 835

\bibitem[{{Mohr} {et~al.}(1999){Mohr}, {Mathiesen}, \& {Evrard}}]{mohr}
{Mohr}, J.~J., {Mathiesen}, B., \& {Evrard}, A.~E. 1999, \apj, 517, 627

\bibitem[{{Molendi} \& {Pizzolato}(2001)}]{molendi01}
{Molendi}, S. \& {Pizzolato}, F. 2001, \apj, 560, 194

\bibitem[{{Mullis} {et~al.}(2004){Mullis}, {Vikhlinin}, {Henry}, {Forman},
  {Gioia}, {Hornstrup}, {Jones}, {McNamara}, \& {Quintana}}]{mullis}
{Mullis}, C.~R., {Vikhlinin}, A., {Henry}, J.~P., {et~al.} 2004, \apj, 607, 175

\bibitem[{{Neumann}(2005)}]{neumann2}
{Neumann}, D.~M. 2005, \aap, 439, 465

\bibitem[{{Neumann} \& {Arnaud}(2001)}]{neumann}
{Neumann}, D.~M. \& {Arnaud}, M. 2001, \aap, 373, L33

\bibitem[{{Ota} {et~al.}(2006){Ota}, {Kitayama}, {Masai}, \& {Mitsuda}}]{ota}
{Ota}, N., {Kitayama}, T., {Masai}, K., \& {Mitsuda}, K. 2006, \apj, 640, 673

\bibitem[{{Peres} {et~al.}(1998){Peres}, {Fabian}, {Edge}, {Allen},
  {Johnstone}, \& {White}}]{peres}
{Peres}, C.~B., {Fabian}, A.~C., {Edge}, A.~C., {et~al.} 1998, \mnras, 298, 416

\bibitem[{{Peterson} {et~al.}(2001){Peterson}, {Paerels}, {Kaastra}, {Arnaud},
  {Reiprich}, {Fabian}, {Mushotzky}, {Jernigan}, \& {Sakelliou}}]{peterson}
{Peterson}, J.~R., {Paerels}, F.~B.~S., {Kaastra}, J.~S., {et~al.} 2001, \aap,
  365, L104

\bibitem[{{Poole} {et~al.}(2008){Poole}, {Babul}, {McCarthy}, {Sanderson}, \&
  {Fardal}}]{poole}
{Poole}, G.~B., {Babul}, A., {McCarthy}, I.~G., {Sanderson}, A.~J.~R., \&
  {Fardal}, M.~A. 2008, \mnras, 391, 1163

\bibitem[{{Poole} {et~al.}(2006){Poole}, {Fardal}, {Babul}, {McCarthy},
  {Quinn}, \& {Wadsley}}]{poole2}
{Poole}, G.~B., {Fardal}, M.~A., {Babul}, A., {et~al.} 2006, \mnras, 373, 881

\bibitem[{{Pratt} {et~al.}(2010){Pratt}, {Arnaud}, {Piffaretti},
  {B{\"o}hringer}, {Ponman}, {Croston}, {Voit}, {Borgani}, \& {Bower}}]{pratt}
{Pratt}, G.~W., {Arnaud}, M., {Piffaretti}, R., {et~al.} 2010, \aap, 511, A85+

\bibitem[{{Predehl} {et~al.}(2010){Predehl}, {Boehringer}, {Brunner}, {Brusa},
  {Burwitz}, {Cappelluti}, {Churazov}, {Dennerl}, {Finoguenov}, {Freyberg},
  {Friedrich}, {Hasinger}, {Kendziorra}, {Kreykenbohm}, {Schmid}, {Wilms},
  {Lamer}, {Meidinger}, {Muehlegger}, {Pavlinsky}, {Robrade}, {Santangelo},
  {Schmitt}, {Schwope}, {Steinmetz}, {Strueder}, {Sunyaev}, \&
  {Tenzer}}]{erosita}
{Predehl}, P., {Boehringer}, H., {Brunner}, H., {et~al.} 2010, ArXiv e-prints

\bibitem[{{Reiprich} \& {B{\"o}hringer}(2002)}]{reip}
{Reiprich}, T.~H. \& {B{\"o}hringer}, H. 2002, \apj, 567, 716

\bibitem[{{Reiprich} {et~al.}(2004){Reiprich}, {Sarazin}, {Kempner}, \&
  {Tittley}}]{reip04}
{Reiprich}, T.~H., {Sarazin}, C.~L., {Kempner}, J.~C., \& {Tittley}, E. 2004,
  \apj, 608, 179

\bibitem[{{Rossetti} \& {Molendi}(2010)}]{rm09}
{Rossetti}, M. \& {Molendi}, S. 2010, \aap, 510, A83+

\bibitem[{{Sakelliou} \& {Ponman}(2004)}]{sakelliou}
{Sakelliou}, I. \& {Ponman}, T.~J. 2004, \mnras, 351, 1439

\bibitem[{{Santos} {et~al.}(2010){Santos}, {Tozzi}, {Rosati}, \&
  {B{\"o}hringer}}]{santos}
{Santos}, J.~S., {Tozzi}, P., {Rosati}, P., \& {B{\"o}hringer}, H. 2010, \aap,
  521, A64+

\bibitem[{{Vikhlinin} {et~al.}(2007){Vikhlinin}, {Burenin}, {Forman}, {Jones},
  {Hornstrup}, {Murray}, \& {Quintana}}]{vikhlinin}
{Vikhlinin}, A., {Burenin}, R., {Forman}, W.~R., {et~al.} 2007, in Heating
  versus Cooling in Galaxies and Clusters of Galaxies, ed. {H.~B{\"o}hringer,
  G.~W.~Pratt, A.~Finoguenov, \& P.~Schuecker }, 48--+

\end{thebibliography}

\appendix

\section{XMM-Newton/MOS2 vs ROSAT/PSPC cross-calibration}
\label{calib}
\normalsize{
The present work offers an excellent opportunity to study the flux cross-calibration between the instruments used (\textit{XMM-Newton}/MOS2 and \textit{ROSAT}/PSPC). Indeed, the HIFLUGCS sample contains bright objects which are intrinsically persistent over very long timescales, unlike the majority of X-ray emitting sources. In this appendix, we present a comparison between the unabsorbed fluxes in the $0-r_{500}$ radial range for 37 of the 64 HIFLUGCS clusters, for which we performed the analysis by combining both instruments. The surface-brightness profiles were fitted simultaneously with the same physical parameters for both instruments, while only the normalizations and background levels were adjusted individually. For the details of the flux reconstruction procedure, see Sect. \ref{fluxes}.}

\begin{figure*}
\resizebox{\hsize}{!}{\mbox{\includegraphics{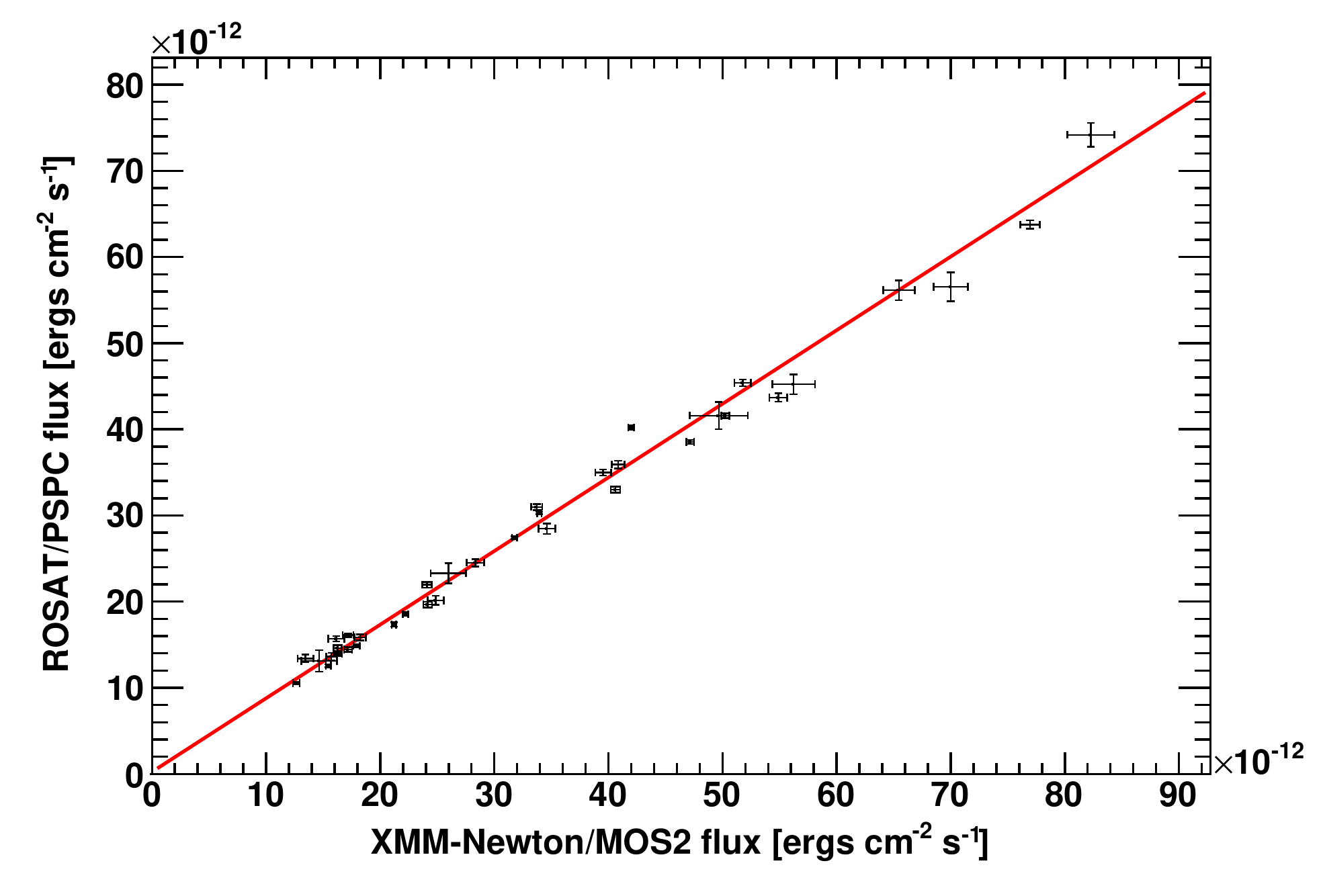}\includegraphics{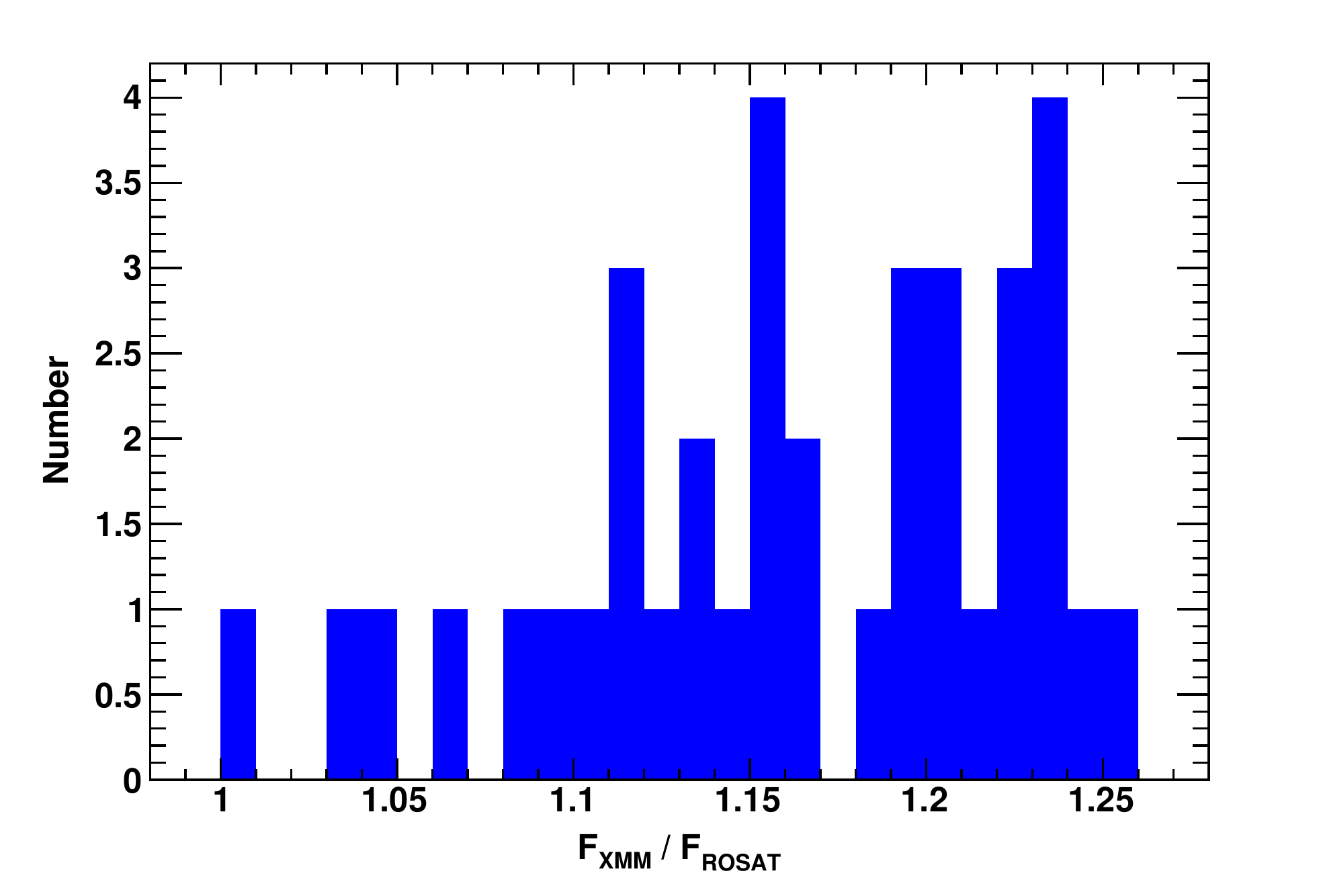}}}
\caption{Comparison between \textit{XMM-Newton}/MOS2 and \textit{ROSAT}/PSPC unabsorbed fluxes in the 0.5-2.0 keV band. Left: PSPC fluxes vs MOS2 fluxes fitted by a simple linear relationship. Right: Ratio between the MOS2 and PSPC fluxes.}
\label{mos2vspspc}
\end{figure*}

Figure \ref{mos2vspspc} shows the comparison between PSPC and MOS2 fluxes in the 0.5-2.0 keV band. The left panel shows the PSPC fluxes as a function of the MOS2 fluxes, fitted by a simple linear relationship. The right panel shows the distribution of the ratio between the MOS2 and PSPC fluxes. The best-fit relation gives

\be F_{ROSAT}=(0.854\pm0.004)F_{XMM}+(2.4\pm1.1)\times10^{-13} ,\ee

\noindent where the fluxes are expressed in units of ergs cm$^{-2}$ s$^{-1}$. The scatter of the relation is 5\%. 

In conclusion, we can see that a systematic difference in absolute calibration is clearly present, \textit{XMM-Newton} giving 15\% higher fluxes with respect to \textit{ROSAT}.

\section{Notes on individual clusters}
\label{indiv}

\begin{itemize}

\item{\large{\emph{A85}}: \normalsize{This cluster is categorized as SCC but appears to be merging with at least one group South of the core \citep{kempner}. This substructure was excised when extracting the surface-brightness profile.\\}}

\item{\large{\emph{NGC 507}}: \normalsize{The surface-brightness profile of this nearby group ($z=0.0165$) is not well represented by either a single beta or a double beta model ($\chi^2=784.5$/261 d.o.f. for the best fit with a single beta model).\\}}

\item{\large{\emph{A399/A401}}: \normalsize{These two clusters appear to be connected \citep{sakelliou}. The \textit{ROSAT}/PSPC pointed observation was pointed on the centre of A401. The surface-brightness profile for A401 was therefore extracted in a sector excluding A399.\\}}

\item{\large{\emph{Fornax}}: \normalsize{This nearby poor cluster shows a main peak on the BCG NGC 1399 and a secondary peak on NGC 1404. NGC 1399 was chosen as the center of the cluster and the area surrounding NGC 1404 was ignored when extracting the surface-brightness profile. Overall, even when using the large \textit{ROSAT} FOV only a small fraction of $r_{500}$ is observed, hence the errors when extrapolating the fluxes can be large. Moreover, the surface-brightness profile of the source is not well fitted by any simple model, so the results of the fitting procedure might be unstable. We consider this object as a very peculiar case.\\}}

\item{\large{\emph{NGC 1550}}: \normalsize{This very nearby group ($z=0.0123$) was observed only by \textit{XMM-Newton}. As a result, less than half of $r_{500}$ was observed within the FOV of the instrument, so the extrapolated fluxes are rather uncertain.\\}}

\item{\large{\emph{A3376}}: \normalsize{A very disturbed cluster which shows an elongated, cometary shape, and two bow-shock-like giant radio relics \citep{bagchi}; no obvious center can be defined. Using the emission centroid as the center, this cluster shows an anomalously high core radius ($r_{c1}=625\pm31$ kpc) and an extremely steep decline ($\beta=1.26\pm0.08$). 
As a result, the model gives only a rough estimate of the surface brightness.\\}}

\item{\large{\emph{A3395s}}: \normalsize{This cluster appears to be merging with a smaller structure in the East \citep[A3395e,][]{donnelly}. The surface-brightness profile was extracted in a sector excluding the secondary structure.\\}}

\item{\large{\emph{A754}}: \normalsize{The morphology of this well-known merging cluster is very disturbed \citep[see][for a detailed analysis of the \textit{XMM-Newton} data]{henry}, so no obvious center can be defined. We used the emission centroid as the center.\\}}

\item{\large{\emph{NGC 4636}}: \normalsize{The surface-brightness profile of this nearby elliptical in the outskirts of the Virgo cluster is not well represented by any simple model ($\chi^2=1951.7$ for 259 d.o.f.). As a result, the output parameters are uncertain.\\}}

\item{\large{\emph{A1644}}: \normalsize{This cluster shows a double-peak structure \citep{reip04}. We extracted the surface-brightness profile in a sector avoiding the North substructure.\\}}

\item{\large{\emph{A1736}}: \normalsize{This is a difficut case, since no \textit{ROSAT}/PSPC observation exists and the available \textit{XMM-Newton} observation was affected by a high background level. As a result, the $\beta$ value could not be constrained, and was fixed to the standard value of $0.67$ while fitting. We note however that our output fluxes are in good agreement with the original flux from \citep{reip}.\\}}

\item{\large{\emph{MKW 8}}: \normalsize{This cluster extends well beyond the FOV of \textit{XMM-Newton}, which introduces rather large errors when extrapolating the fluxes to $r_{500}$.\\}}

\item{\large{\emph{A2634}}: \normalsize{The \textit{ROSAT}/PSPC image shows a secondary peak North-West of the main cluster. This structure was excised for the extraction of the surface-brightness profile.\\}}

\item{\large{\emph{A4038}}: \normalsize{It is not clear whether the surface-brightness profile of this cluster should be modeled with a single beta or a double beta. A better fit is achieved when using a double beta model, but the improvement to the fit is not significant, so the single beta model was used.\\}}

\end{itemize}

\end{document}